\documentclass[12pt,preprint]{aastex}
\usepackage{lscape,graphicx}
\usepackage{rotating}

\newcommand{\kms}{km s$^{-1}$}
\newcommand{\hh}{H$_2$}
\newcommand{\ceo}{C$^{18}$O}

\shorttitle{Molecular Tracers of Embedded Star Formation}
\shortauthors{Gurney et al.}

\begin{document}

\title{Molecular Tracers of Embedded Star Formation in Ophiuchus}

\author{M. Gurney\altaffilmark{1}, R. Plume\altaffilmark{1,2}, D. Johnstone\altaffilmark{3}}

\altaffiltext{1}{Centre for Radio Astronomy, University of Calgary, 2500 University Dr. NW, Calgary, AB, T2N 1N4, Canada, mgurney@phas.ucalgary.ca; plume@ism.ucalgary.ca}
\altaffiltext{2}{Max Planck Institute for Astronomy, K\"onigstuhl 17, 69117 Heidelberg, Germany, plume@ism.ucalgary.ca}
\altaffiltext{3}{NRC-Herzberg Institute of Astrophysics, 5071 W. Saanich Road, Victoria, BC, V9E 2E7,  Canada, Douglas.Johnstone@nrc-cnrc.gc.ca}

\newpage

\begin{abstract}

In this paper we analyze nine SCUBA cores in Ophiuchus using the second-lowest rotational transitions of 
four molecular species ($^{12}$CO, $^{13}$CO, C$^{18}$O, and C$^{17}$O) to search for clues to the evolutionary state and star-formation activity within each core.
Specifically, we look for evidence of outflows, infall, and CO depletion. The line wings in the CO spectra
are used to detect outflows, spectral asymmetries in $^{13}$CO are used to determine infall characteristics, and 
a comparison of the dust emission (from SCUBA observations) and gas emission (from C$^{18}$O) is used to determine
the fractional CO freeze-out. 

Through comparison with Spitzer observations of protostellar sources in Ophiuchus, 
we discuss the usefulness of CO and its 
isotopologues as the sole indicators of the evolutionary state of each core.  This study is an important pilot project for 
the JCMT Legacy Survey of the Gould Belt (GBS) and the Galactic Plane (JPS), which intend to complement the SCUBA-2 
dust continuum observations with HARP observations of $^{12}$CO, $^{13}$CO, C$^{18}$O, and C$^{17}$O J = 3$\rightarrow$2 in
order to determine whether or not the cold dust clumps detected by SCUBA-2 are protostellar or starless objects.

Our classification of the evolutionary state of the cores (based on molecular line maps and SCUBA observations) is in agreement with the Spitzer designation for six or seven of the nine SCUBA cores.  However, several important caveats must be noted in the interpretation of these results.  First, while these tracers may work well in isolated cores, care must be taken in blindly applying these metrics to crowded regions. Maps of larger areas at higher resolution are required to determine whether the detected outflows originate from the core of interest, or from an adjacent core with an embedded YSO.  Second, the infall parameter may not be an accurate tracer of star-formation activity because global motions of the cloud may act to emulate what appears to be the collapse of a single core.

Large mapping surveys like the GBS may be able to overcome some of this confusion and disentangle one outflow from another by mapping the full extent of the outflows and allowing us to find the originating object. As well, the higher-energy CO J = $3\rightarrow2$ transition used by the GBS has a higher critical density and so will trace the warm dense gas in the outflow rather than the lower density surrounding cloud material. The higher resolution of the GBS observations at 345 GHz ($\theta_{FWHM} \approx 14''$ vs. our $22''$) may also provide a clearer picture of activity in crowded fields.

\end{abstract}


\section{Introduction}
\label{sec:intro}

There is a strong correlation between star formation and structure within molecular clouds, with
young stars forming in the densest cores. Recent large-coverage submillimeter continuum maps of 
star-forming molecular clouds have unveiled hundreds of previously unidentified dense cores and 
it is intriguing to ask whether the majority of these cores are, or will, convert themselves into 
stars. Answering this question requires evidence of the existence, or absence, of embedded protostars, and a knowledge of
their evolutionary state.  The Spitzer Space Telescope, imaging in the mid-infrared, can reveal the 
locations of protostars,  but their evolutionary state needs to be disentangled
either from the amount of gas available in the core versus the protostellar mass, an almost 
impossible task at present, or from some other evolutionary diagnostic such as jet strength and
morphology.

There has been a lot of recent work connecting the core mass function (CMF) with the stellar initial mass function (IMF). 
These mass functions are most often parameterized as
${{dN}\over{dM}} \propto M^{-\alpha}$, with the IMF having
$\alpha \approx  2.35$ for 0.5\,M$_\odot <$ M $<$ 10\,M$_\odot$ (e.g. Salpeter 1955), or more recently,  
$\alpha \approx 2.2$ for M $> 0.5$M$_\odot$ and $\alpha \approx 1.3 \pm 0.5 $ for M $< 0.5 $M$_\odot$ (Kroupa 2001 and references therein). Motte et al. (1998) undertook 
a 1.3\,mm continuum survey in Ophiuchus and found the core mass spectrum to have 
$\alpha \approx 2.5$ for 0.5\,M$_{\odot} < $ M $< $  3\,M$_\odot$ and 
$\alpha \approx 1.5$ for 0.1\,M$_{\odot} <$ M $< $ 0.5\,M$_\odot$. Testi \& Sargent (1998) found 
similar results in Serpens with $\alpha \approx$ 2.1 for M $ > 0.3\,$M$_\odot$. Further work by
Johnstone et al. (2000, 2001) in the Ophiuchus and Orion molecular clouds, and Lada et al. (2008)
in the Pipe Nebula have found similar core mass functions. Johnstone et al. (2000), however, suggest that 
the cores identified in these surveys may not be collapsing, and that the majority of the cores, 
when modeled as Bonnor-Ebert spheres, are not dominated by gravitational forces and, thus, should only
collapse if there is a significant increase in the external pressure or if the internal temperature 
drops. For Bonnor-Ebert spheres, there exists a maximum central density enhancement, due to 
gravity, for which stable structures are possible. For an observed surface density profile 
the central density enhancement can be written in terms of a concentration parameter, C, defined as  
 
 \begin{equation}
 C = 1 - \frac{M}{\pi R_{eff}^2 \Sigma_o}
 \end{equation}\\
 where M is the mass of the core, R$_{eff}$ is its effective radius, $\Sigma_o$ is the central column density (Johnstone et al. 2000). C $>$ 0.72 means that thermal pressure alone cannot support the structure, and magnetic fields or turbulence will be required if collapse is to be prevented. C $<$ 0.72 means that the Bonnor-Ebert sphere will not collapse under gravity. In the Perseus molecular cloud (J\o rgensen et al. 2007) the concentration parameter was found to be a sufficient diagnostic for embedded stars but not a
necessary one. 

The ongoing James Clerk Maxwell Telescope (JCMT) Legacy Gould Belt Survey (GBS; Ward-Thompson et al. 2007) will further investigate 
our understanding of how the core mass spectrum relates to the IMF.  The GBS will observe several low- 
and intermediate-mass star-forming regions in the Gould Belt in both submillimeter dust continuum and molecular lines. 
The Gould Belt is a ring of O-type stars which is inclined 20$^\circ$ to the galactic plane, is centered 
at a point $\approx$ 200\,pc from the Sun, and is about 350\,pc in radius. It is a highly active star-forming 
ring and contains many well-known star-forming regions including Ophiuchus, Orion, Serpens, Perseus, Taurus, Auriga, 
and Scorpius (Clube 1967; Stothers \& Frogel 1974; Comeron et al. 1992). The GBS will survey over 700  deg$^2$ of sky with the Submillimetre Common-User Bolometer Array (SCUBA-2; Ward-Thompson et al. 2007) at 450 and 850$\mu$m,
increasing the number of known starless and protostellar cores by over an order of magnitude and 
establishing the first unbiased catalog of nearby submillimeter continuum sources. The survey will  
also observe a sample of these submillimeter cores in CO and its isotopologues using the Heterodyne Array Receiver Program (HARP; Smith et al. 2003, Buckle et al. 2006). Together, these observations will attempt to identify prestellar and 
protostellar cores through three mechanisms:  outflows, infall kinematics, and column densities. These mechanisms are useful in tracing protostellar evolution, as objects throughout their various stages of evolution will exhibit different strengths of each.  Pre-protostellar objects are characterized by strong infall and depletion, but since no young stellar object (YSO) has yet formed, there is no outflow. Protostellar objects have strong outflows, but their rate of infall and amount of depletion begins to decrease with the formation of a hot young stellar object.  Therefore these objects may or may not exhibit evidence of infall and depletion.

In this paper we present analysis of data  taken in the Ophiuchus molecular cloud,
a close (approximately 160\,pc from Earth), extensively studied low-mass star-forming region in the Gould
Belt, which is approximately 1\,pc$^2$ in size and is made up of several regions, including L1688. L1688 
(also known as the Ophiuchus main cloud) is itself very clumpy and contains a large concentration of newly 
formed stars. The star-formation efficiency in L1688 was found to be about 10\% (J\o rgensen et al. 2008) and about 2.5\% in a larger region of the Ophiuchus cloud (Johnstone et al. 2004).  
Submillimeter observations have revealed over 50 submillimeter cores (Andr\'{e} et al. 1993; Young et al. 2006) with densities of
n $>$ 10$^5$ cm$^{-3}$ and temperatures between 15 and 35K.  The high density of the cores and high star-formation efficiency  make this an ideal region in which to examine the difference between starless and protostellar cores.  
In this study we are aided by the recent work  of  J\o rgensen et al. (2008), which compares data from the  c2d project of the Spitzer Space Telescope Legacy Survey (Evans et al. 2003)  and SCUBA submillimeter continuum data (J\o rgensen et al. 2008), linking individual cores with embedded protostars.

This paper is a pilot project for the JCMT Legacy Gould Belt Survey in which we attempt to determine if 
observations of CO and its isotopologues are accurate tracers of star-formation activity and/or evolution.  
We accomplish this by observing nine SCUBA cores from the Johnstone et al. (2000) survey that have been 
identified as either protostellar or starless via Spitzer observations.  Using the same molecular species as will be used by the GBS, we determine outflow characteristics, infall signatures, and depletion 
factors for each core and compare the results with the Spitzer starless/protostellar classification.

\section{Observations}
\label{sec:obs}

\begin{figure}[h] 
   \centering
 \includegraphics[angle=270, width=6.5in]{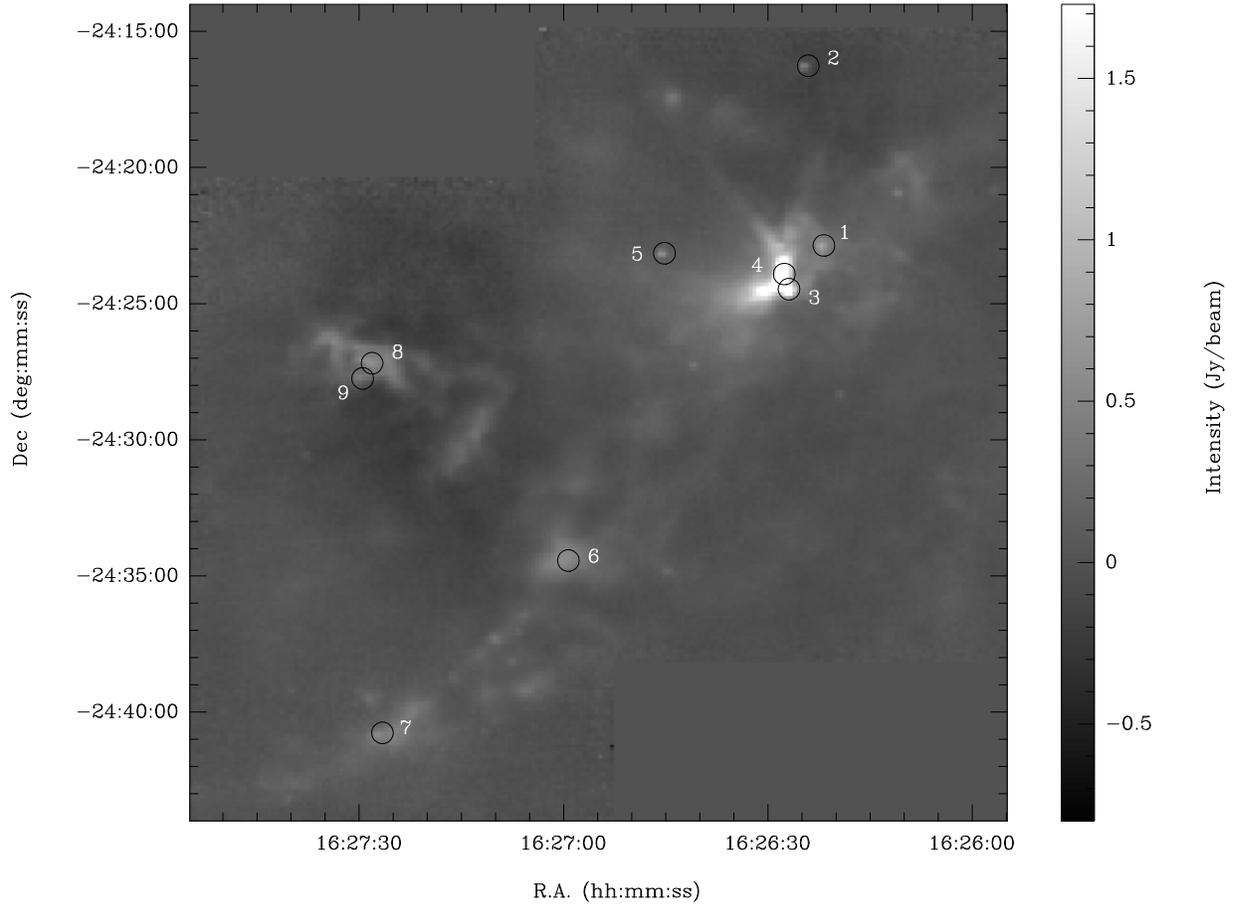} 
   \caption{SCUBA  map of the Ophiuchus region at 850 $\mu$m taken from Johnstone et al. (2000). The beam half-power beamwidth HPBW is 14$''$. The overlaid circles with associated numbers mark the positions of the nine SCUBA cores studied in this paper.} 
   \label{fig:oph_map}
\end{figure}

The sources chosen for this study were selected from a list of cores in Ophiuchus observed at 850 $\mu$m 
by Johnstone et al. (2000). There were only 13 cores in this list with a peak flux greater than 
400 mJy beam$^{-1}$, of which nine were randomly chosen for this study. All of these cores have associated Spitzer 
observations (J\o rgensen et al. 2008). Three of the cores (Sources 2, 5, \& 7) also have coincident Infrared Astronomy Satellite (IRAS) sources. Table \ref{tab:properties} provides the source list, coordinates, designation, concentration, and peak flux at 850 $\mu$m of each SCUBA core (as taken from J\o rgensen et al. 2008), as well as the Johnstone et al. (2000) designations of the cores to which the CO observations were gridded. The uncertainty in the SCUBA flux is about 20\%. Figure \ref{fig:oph_map} shows the 850 $\mu$m maps of the Ophiuchus region mapped by Johnstone et al. (2000) on which we have overlaid the positions of each of our cores.

The Spitzer Space Telescope data for our nine cores were taken as part of the ``From Molecular Cores to 
Planet Forming Disks" Legacy program (also known as``c2d") which aims to study the process of star formation 
from the earliest stages to the point at which planet-forming disks form (Evans et al. 2003). The data were 
taken with the Infrared Array Camera (IRAC), which is composed of four channels that simultaneously provide 
5.2$'\times5.2'$ images at 3.6, 4.5, 5.8, and 8.0 $\mu$m, and the Multiband Imaging Photometer (MIPS), which 
operates at 24, 70, and 160 $\mu$m (Werner et al. 2004). For this analysis we utilized the comparison
between the Spitzer and SCUBA datasets presented by  J\o rgensen et al. (2008).

We mapped the nine cores in Ophiuchus in the CO J = 2$\rightarrow$1 transition at a frequency of 
230.542 GHz, the $^{13}$CO J = 2$\rightarrow$1 transition at a frequency of 220.822 GHz, the C$^{18}$O 
J = 2$\rightarrow$1 transition at 219.146 GHz, and in some of the cores, the C$^{17}$O 
J = 2$\rightarrow$1 transition at a frequency of 224.719 GHz. The spectral resolution for all transitions was 0.2 \kms . These observations were taken in 2003 between February 
and June using the A3 heterodyne receiver at the JCMT.  The $\theta_{FWHM}$ of the CO observations was $\approx 22''$. The average 
system temperature during the observations ranged from $\approx$ 270K to 500K. The atmospheric opacity at 
225 GHz, as measured by the CSO tipping meter was never worse than 0.1.  
40$'' \times 40''$ maps were produced for the 
C$^{17}$O, C$^{18}$O, and $^{13}$CO lines, and 20$'' \times 20''$ maps were produced for the CO lines, all 
with 10$''$ spacing.

All intensity units in this paper are presented in units of T$_A^*$.  To calculate column densities we 
convert to the radiation temperature, T$_R^*$, scale in which T$_R^*$ = T$_A^*$/$\eta_{\rm mb}$ 
(Kutner \& Ulich 1981), where T$_R^*$ is the Rayleigh-Jeans temperature of a spatially resolved source 
observed with a perfect telescope above the atmosphere.  At 230 GHz the main beam efficiency of the JCMT $\eta_{\rm mb}$
is $\approx$ 0.69.

\section{Results}
\label{sec:results}

\begin{figure}[h] 
   \centering
 \includegraphics[angle=270, width=5in]{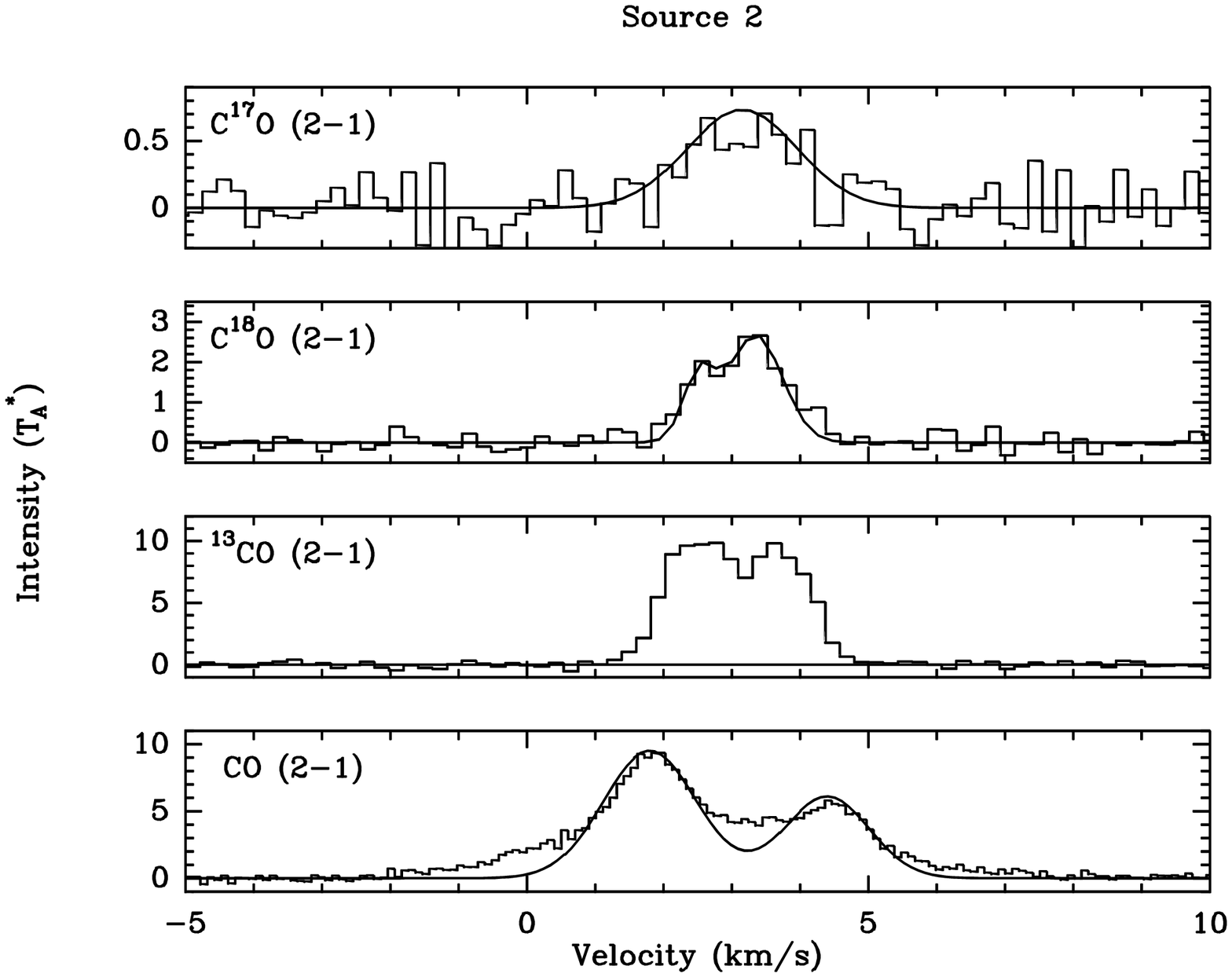} 
   \caption{From top to bottom: C$^{17}$O, C$^{18}$O, $^{13}$CO, and CO J = $2\rightarrow1$ emission from the central position in Source 2. The continuous curves in some of the panels represent single or multiple Gaussian profiles that were fit to the observed data, the results of which were used to determine the C$^{18}$O centroid velocity, integrated intensity, and column density, as well as the extent of the CO line wings (see Table \ref{tab:colden}).}
   \label{fig:Source2_spec}
\end{figure}

\begin{figure}[h] 
   \centering
 \includegraphics[angle=270, width=5in]{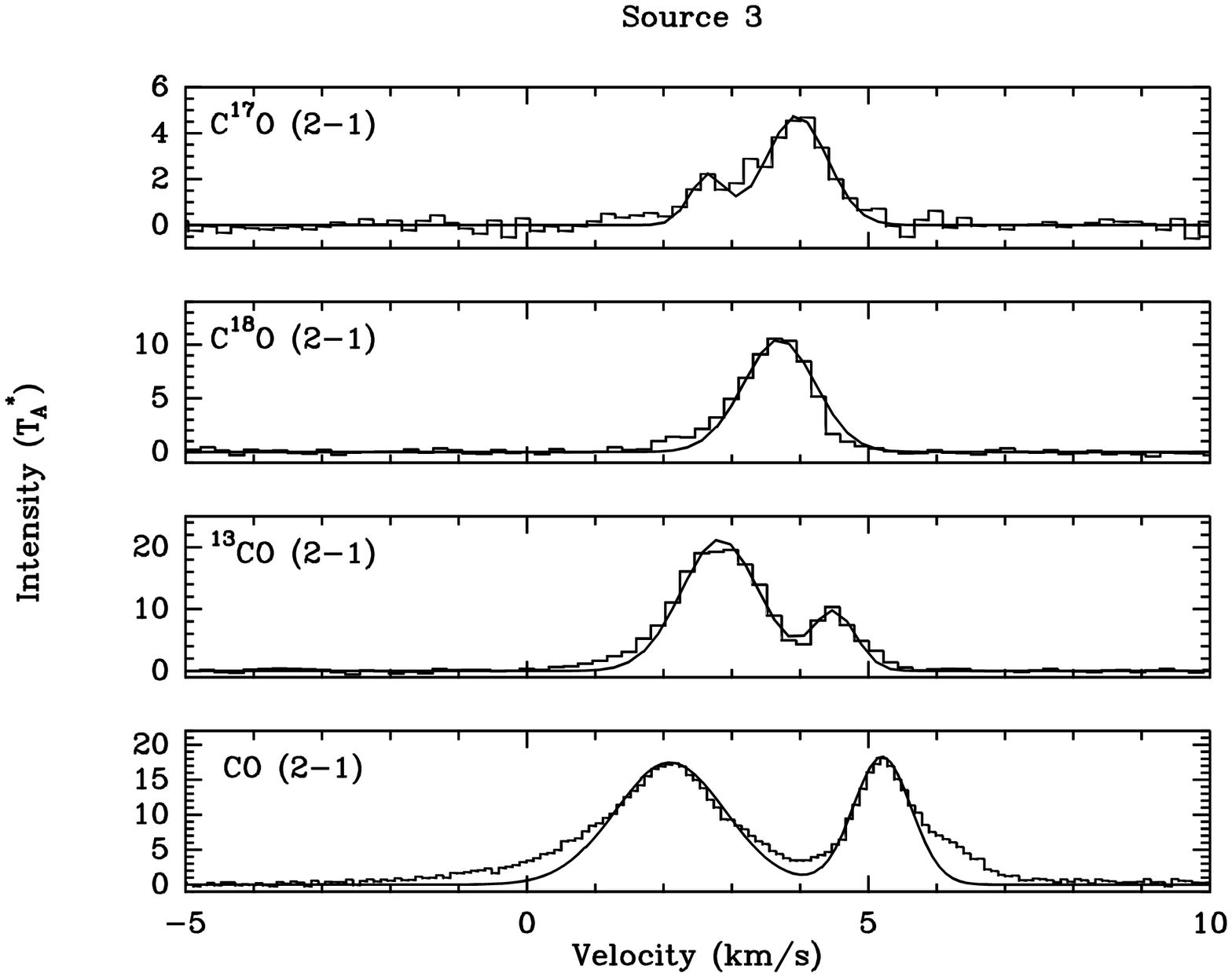} 
   \caption{From top to bottom: C$^{17}$O, C$^{18}$O, $^{13}$CO, and CO J = $2\rightarrow1$ emission from the central position in Source 3. The continuous curves in some of the panels represent single or multiple Gaussian profiles that were fit to the observed data, the results of which were used to determine the C$^{18}$O centroid velocity, integrated intensity, and column density, as well as the extent of the CO line wings (see Table \ref{tab:colden}).}
   \label{fig:Source3_spec}
\end{figure}

\begin{figure}[h] 
   \centering
 \includegraphics[angle=270, width=5in]{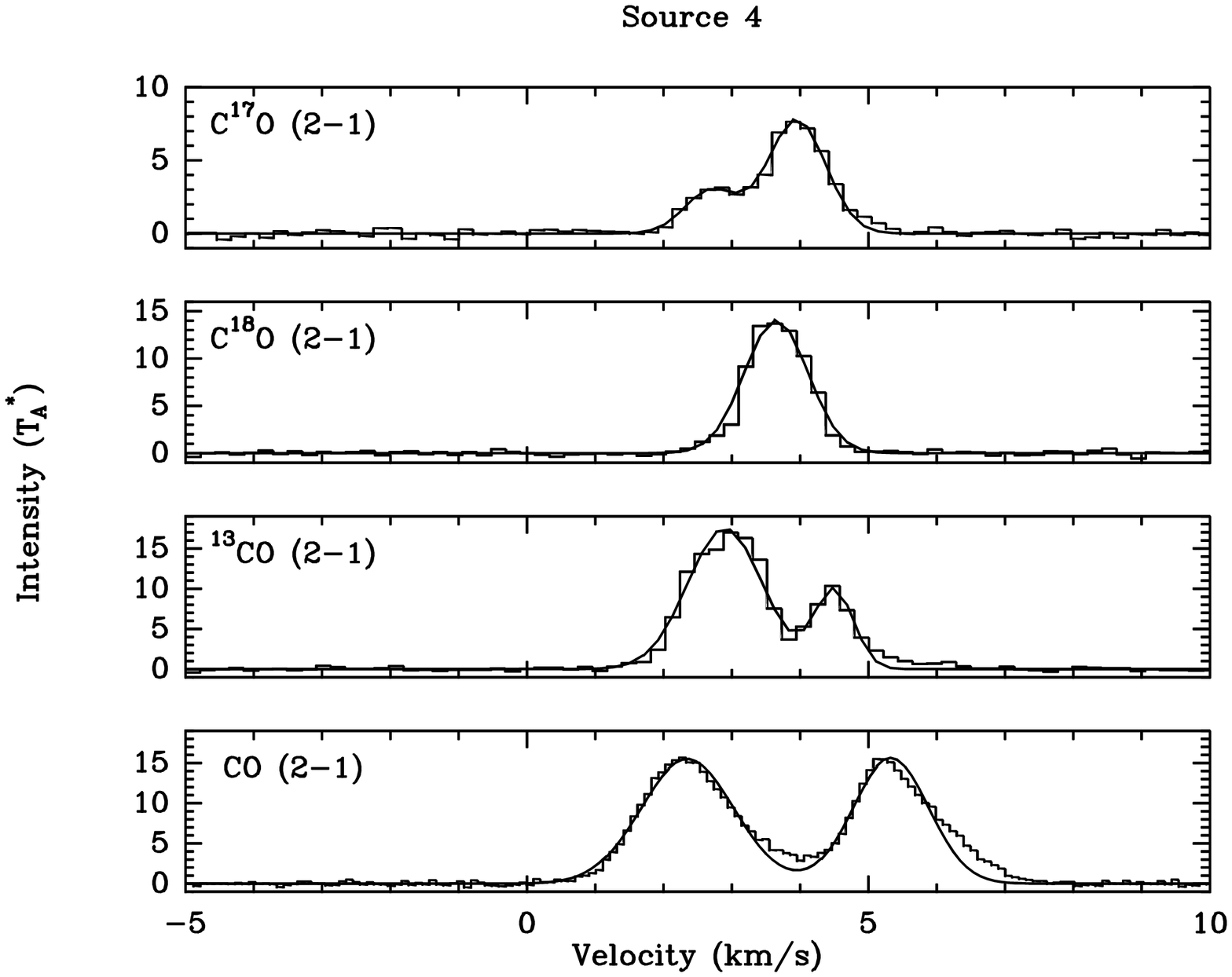} 
   \caption{From top to bottom: C$^{17}$O, C$^{18}$O, $^{13}$CO, and CO J = $2\rightarrow1$ emission from the central position in Source 4. The continuous curves in some of the panels represent single or multiple Gaussian profiles that were fit to the observed data, the results of which were used to determine the C$^{18}$O centroid velocity, integrated intensity, and column density, as well as the extent of the CO line wings (see Table \ref{tab:colden}).}
   \label{fig:Source4_spec}
\end{figure}

\begin{figure}[h] 
   \centering
 \includegraphics[angle=270, width=5in]{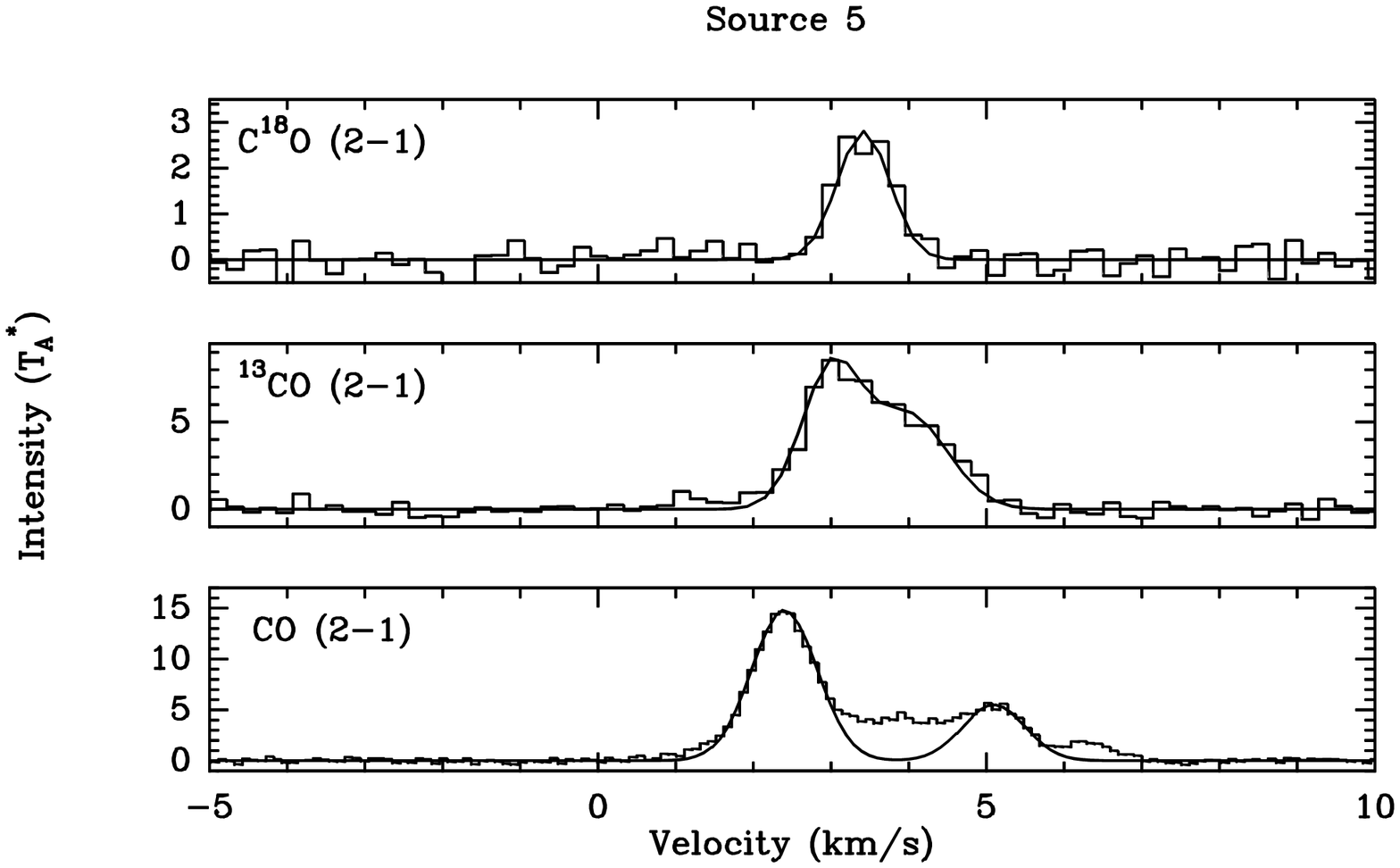} 
   \caption{From top to bottom: C$^{18}$O, $^{13}$CO, and CO J = $2\rightarrow1$ emission from the central position in Source 5. The continuous curves in some of the panels represent single or multiple Gaussian profiles that were fit to the observed data, the results of which were used to determine the C$^{18}$O centroid velocity, integrated intensity, and column density, as well as the extent of the CO line wings (see Table \ref{tab:colden}).}
   \label{fig:Source5_spec}
\end{figure}

\begin{figure}[h] 
   \centering
 \includegraphics[angle=270, width=5in]{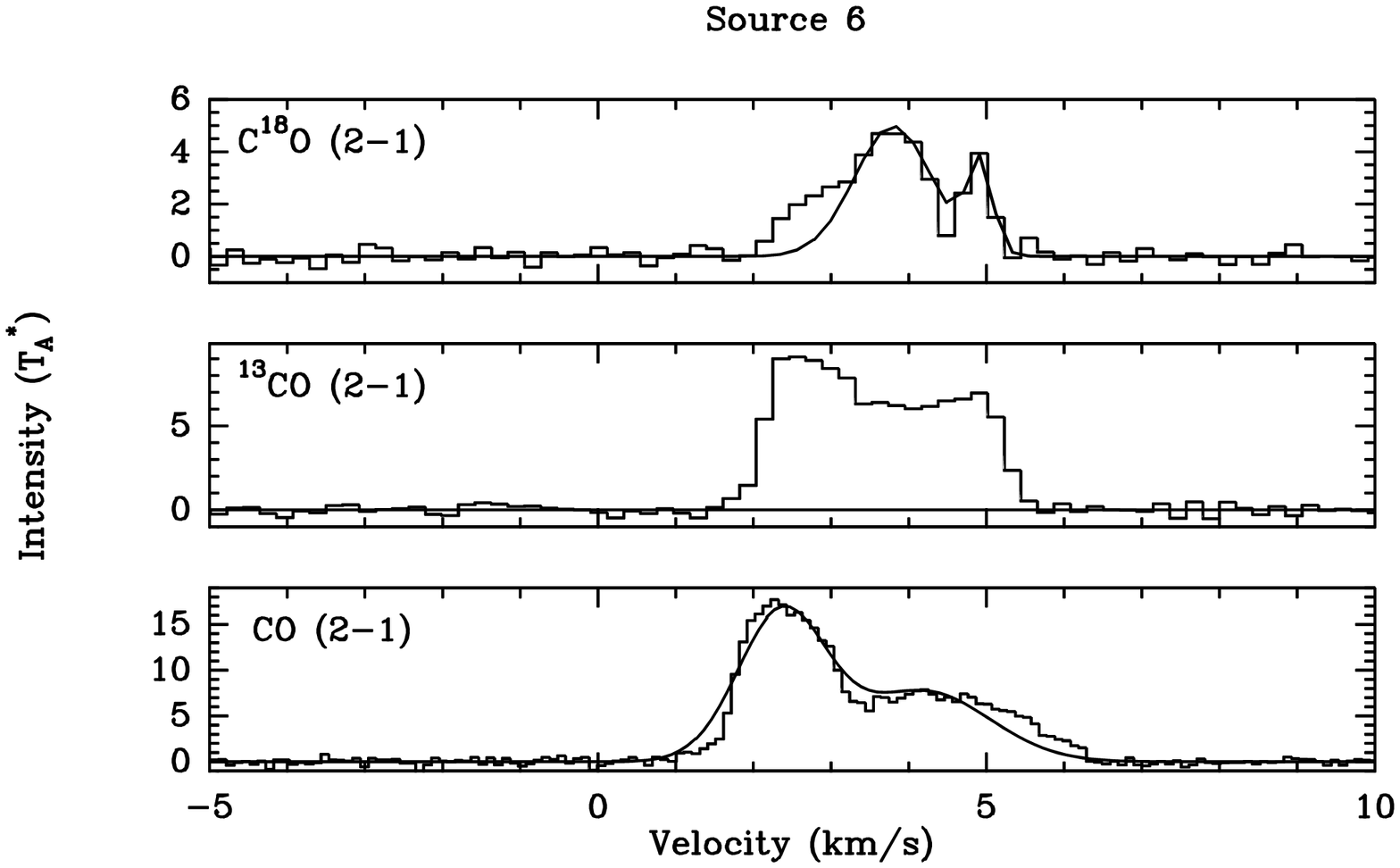} 
   \caption{From top to bottom: C$^{18}$O, $^{13}$CO, and CO J = $2\rightarrow1$ emission from the central position in Source 6. The continuous curves in some of the panels represent single or multiple Gaussian profiles that were fit to the observed data, the results of which were used to determine the C$^{18}$O centroid velocity, integrated intensity, and column density, as well as the extent of the CO line wings (see Table \ref{tab:colden}).}
   \label{fig:Source6_spec}
\end{figure}

\begin{figure}[h] 
   \centering
 \includegraphics[angle=270, width=5in]{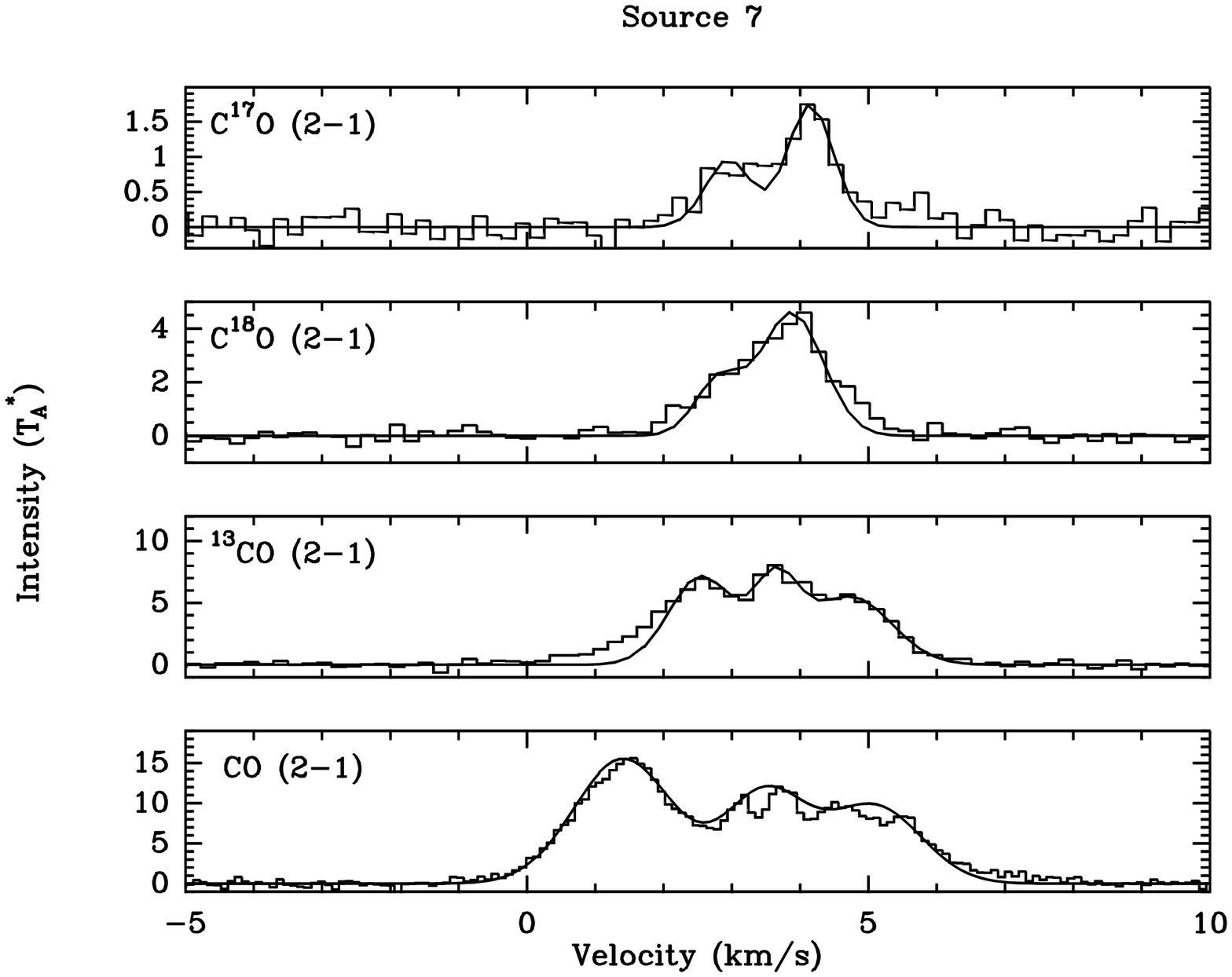} 
   \caption{From top to bottom: C$^{17}$O, C$^{18}$O, $^{13}$CO, and CO J = $2\rightarrow1$ emission from the central position in Source 7. The continuous curves in some of the panels represent single or multiple Gaussian profiles that were fit to the observed data, the results of which were used to determine the C$^{18}$O centroid velocity, integrated intensity, and column density, as well as the extent of the CO line wings (see Table \ref{tab:colden}).}
   \label{fig:Source7_spec}
\end{figure}

\begin{figure}[h] 
   \centering
 \includegraphics[angle=270, width=5in]{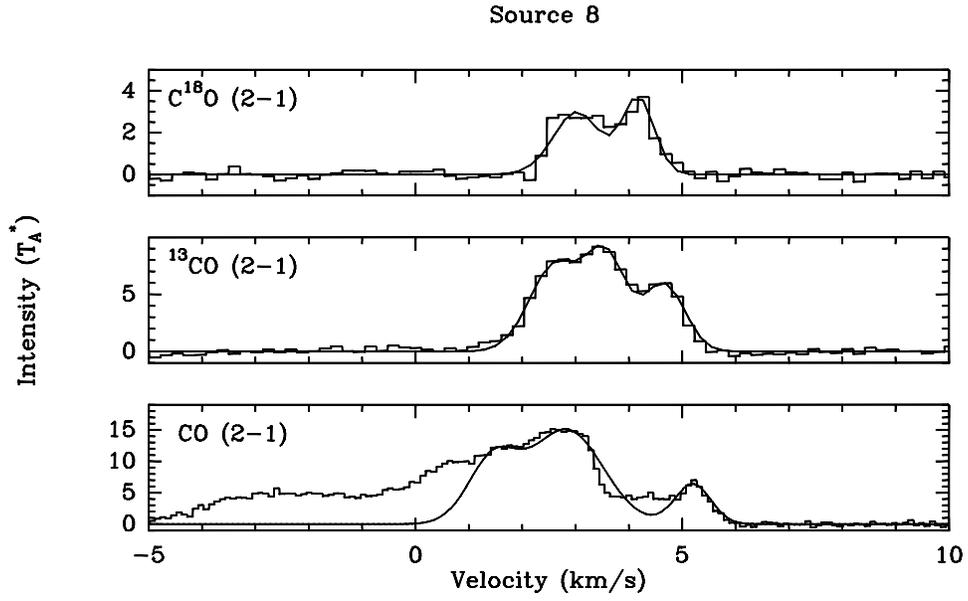} 
   \caption{From top to bottom: C$^{18}$O, $^{13}$CO, and CO J = $2\rightarrow1$ emission from the central position in Source 8. The continuous curves in some of the panels represent single or multiple Gaussian profiles that were fit to the observed data, the results of which are presented in Table \ref{tab:colden}.}
   \label{fig:Source8_spec}
\end{figure}

\begin{figure}[h] 
   \centering
 \includegraphics[angle=270, width=5in]{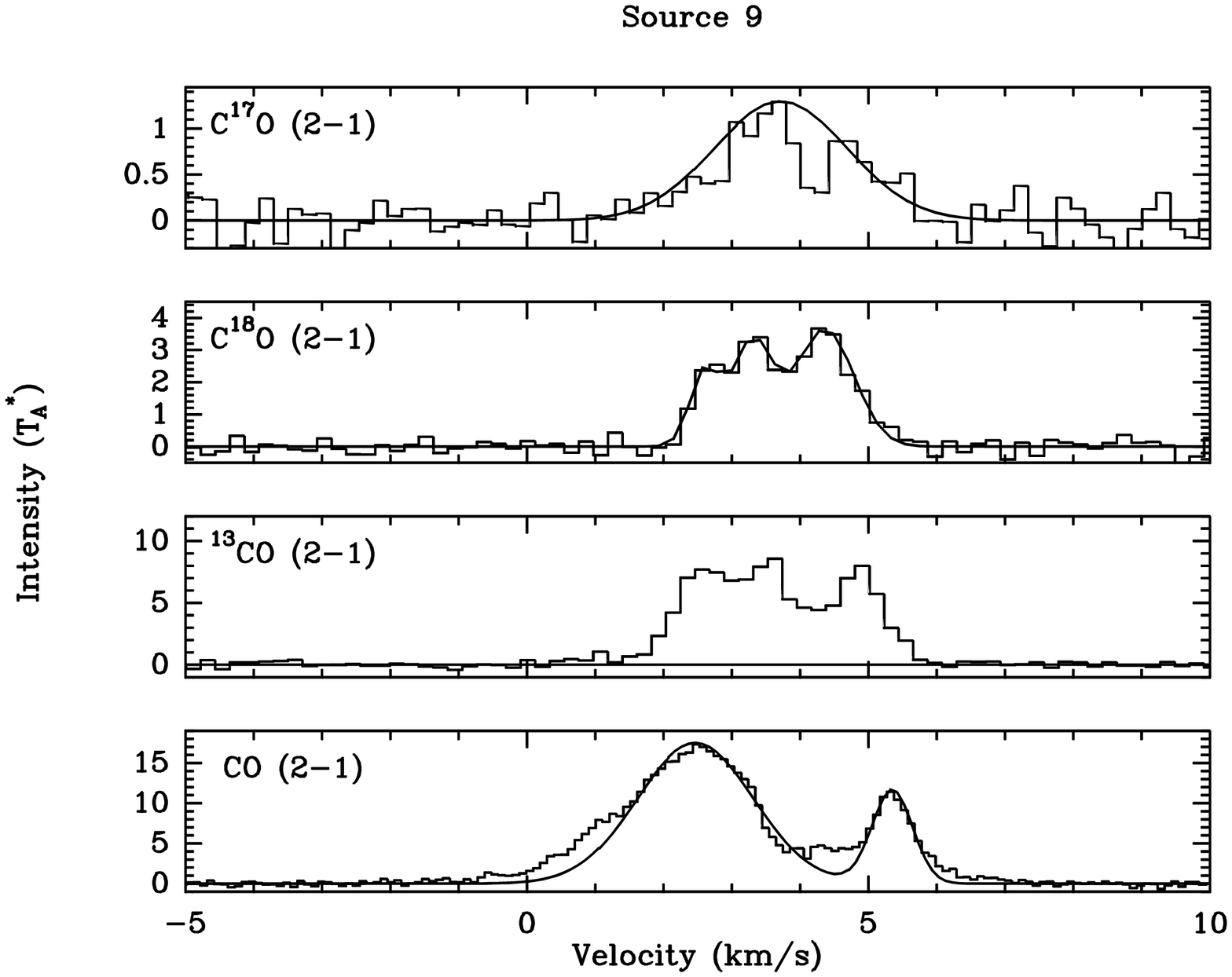} 
   \caption{From top to bottom: C$^{17}$O, C$^{18}$O, $^{13}$CO, and CO J = $2\rightarrow1$ emission from the central position in Source 9. The continuous curves in some of the panels represent single or multiple Gaussian profiles that were fit to the observed data, the results of which were used to determine the C$^{18}$O centroid velocity, integrated intensity, and column density, as well as the extent of the CO line wings (see Table \ref{tab:colden}).}
   \label{fig:Source9_spec}
\end{figure}

In order to determine the usefulness of CO and its isotopologues as a tracer of star-formation
activity in, and evolution of,  cold dust clumps identified by submillimeter continuum observations, we examine 
the molecular line observations for certain key characteristics: evidence for molecular outflows,
evidence for infall, evidence for CO depletion.  These characteristics together are used to identify the evolutionary 
state of each core and the results of this analysis are compared against the Spitzer classification of the cores.
Figures \ref{fig:Source1_spec} to \ref{fig:Source9_spec} show the spectra in each CO isotopologue toward the center
of each core. The histogram-like curves in these figures represent the observed spectra, while the continuous curves in some of the panels represent single or multiple component Gaussian profiles that were fitted to the observed data through the CLASS Gaussian fitting routine, which uses an iterative procedure to find the best fits. The number of components to be fit in each spectra were determined through visual examination of the spectra. The dashed line on the CO emission in Source 1 is an LTE model of an optically thick CO line.

The CO spectra in all sources are extremely self-absorbed, often showing several self-absorption dips 
(e.g., Source 7). The CO spectra also show clear line wings in many cases (e.g., Source 1). The slightly thinner 
$^{13}$CO line also exhibits some self-absorption, often with a blue skewed asymmetric profile. In some cores 
this is shown clearly (e.g., Source 4), while in others the self-absorption profile is confused with other 
spectral components (e.g., Source 6).

Only four of the cores (Sources 1, 3, 4, \& 5) have spectral profiles indicative of simple kinematics:  
a single velocity component in the optically thin C$^{18}$O.  The others show complicated motions
in the vicinity of the cores, displaying several spectral features even in C$^{18}$O, indicating multiple velocity components along the line of sight.

The most optically thin CO isotopologue C$^{17}$O was observed in two-thirds of the sample (Sources 1, 2, 3, 4, 7, \& 9).  C$^{17}$O spectra contain hyperfine structure, determined through laboratory 
measurements to have a separation of 1.120 MHz corresponding to approximately 1.5 km s$^{-1}$ 
(Klapper et al. 2003). This calculation has an uncertainty of about 0.1 km s$^{-1}$. The second spectral features in Sources 1, 3, \& 4 are separated from their main peak by approximately 1.5 km s$^{-1}$ within these uncertainties. Note that the offset between the C$^{17}$O main peak and the centroid of the C$^{18}$O line is likely also due to the uncertainty in the measured C$^{17}$O frequency (Klapper et al. 2003).

\begin{figure}[h] 
   \centering
 \includegraphics[angle=270, width=5in]{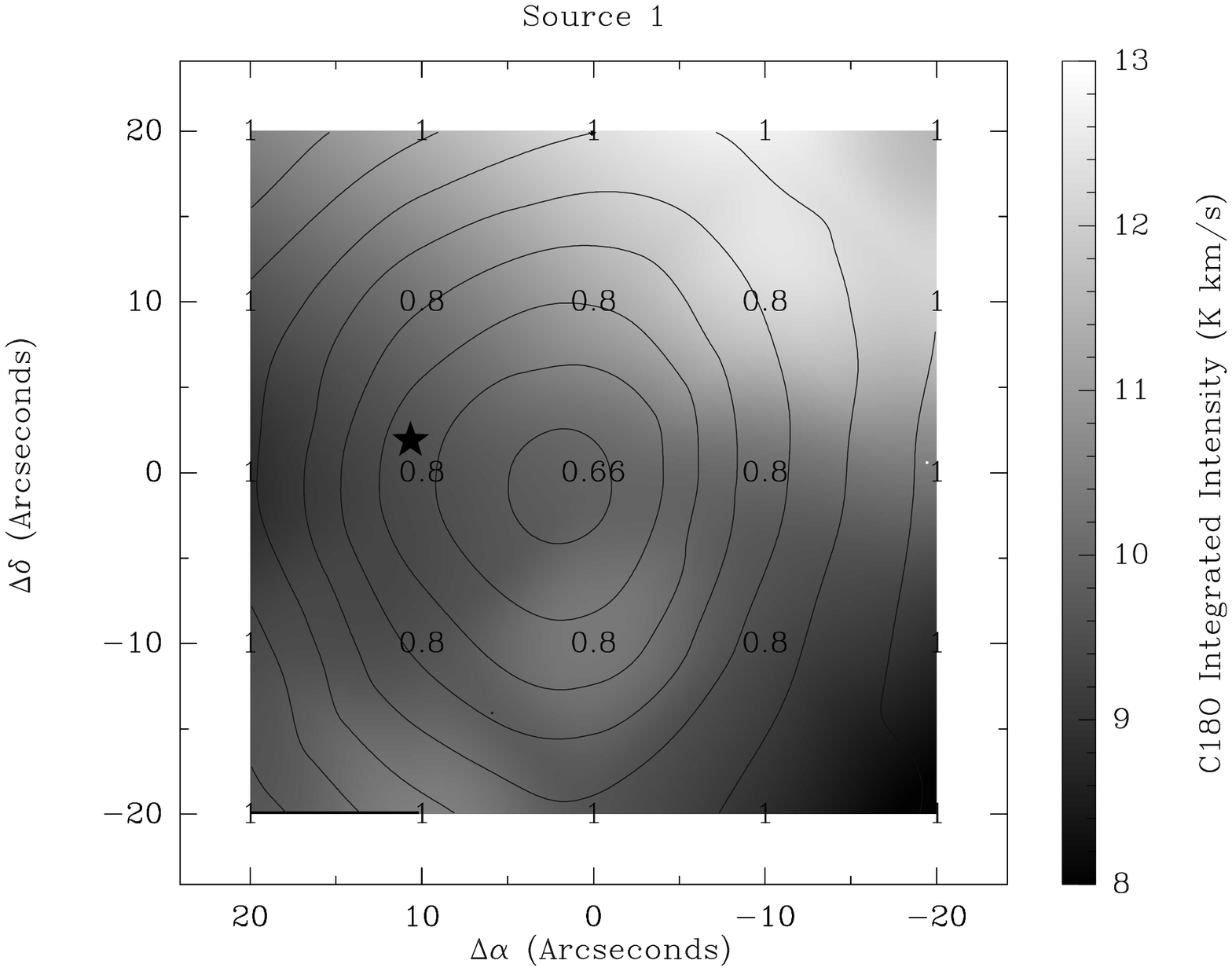} 
   \caption{Contour images of 850 $\mu$m continuum from SCUBA (in units of Jy beam$^{-1}$) smoothed to the CO beamsize and superimposed on C$^{18}$O integrated intensity maps (gray scale) in units of  $\int{T_A^*dV}$ for Source 1. Contour levels are from 0.64 Jy beam$^{-1}$ to 1.28 Jy beam$^{-1}$ in steps of 0.08 Jy beam$^{-1}$. The beam HPBW of the C$^{18}$O observations is $\approx$ 22.8$''$. The numbers indicate the ratio of column density calculated from the C$^{18}$O to column density calculated from the dust. The star marks the position of the Spitzer source.} 
   \label{fig:Source1_con}
\end{figure}

\begin{figure}[h] 
   \centering
 \includegraphics[angle=270, width=5in]{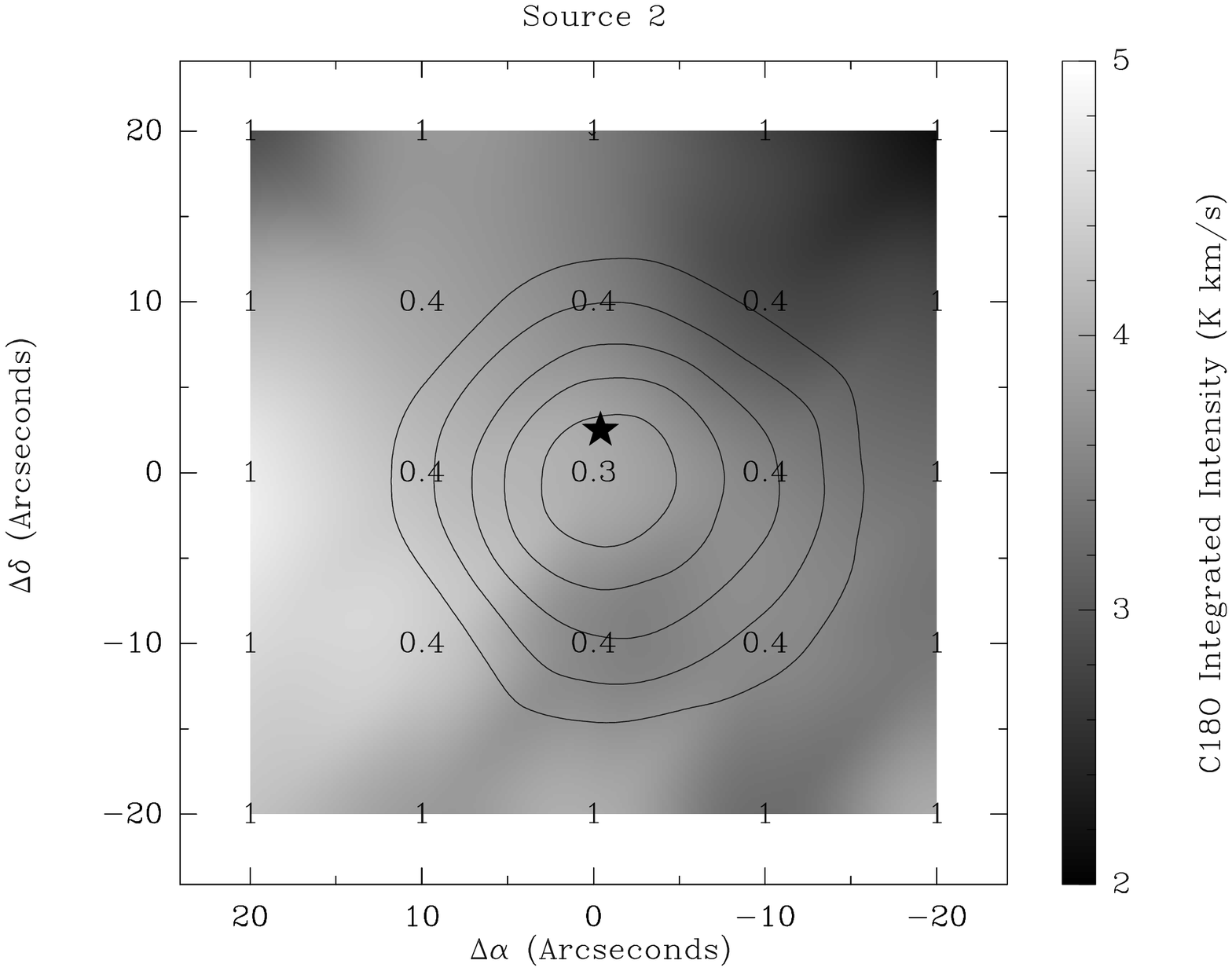} 
   \caption{Contour images of 850 $\mu$m continuum from SCUBA (in units of Jy beam$^{-1}$) smoothed to the CO beamsize and superimposed on C$^{18}$O integrated intensity maps (gray scale) in units of  $\int{T_A^*dV}$ for Source 2. Contour levels are from 0.01 Jy beam$^{-1}$ to 0.32 Jy beam$^{-1}$ in steps of 0.08 Jy beam$^{-1}$. The beam HPBW of the C$^{18}$O observations is $\approx$ 22.8$''$. The numbers indicate the ratio of column density calculated from the C$^{18}$O to column density calculated from the dust. The star marks the position of the Spitzer source.} 
   \label{fig:Source2_con}
\end{figure}

\begin{figure}[h] 
   \centering
 \includegraphics[angle=270, width=5in]{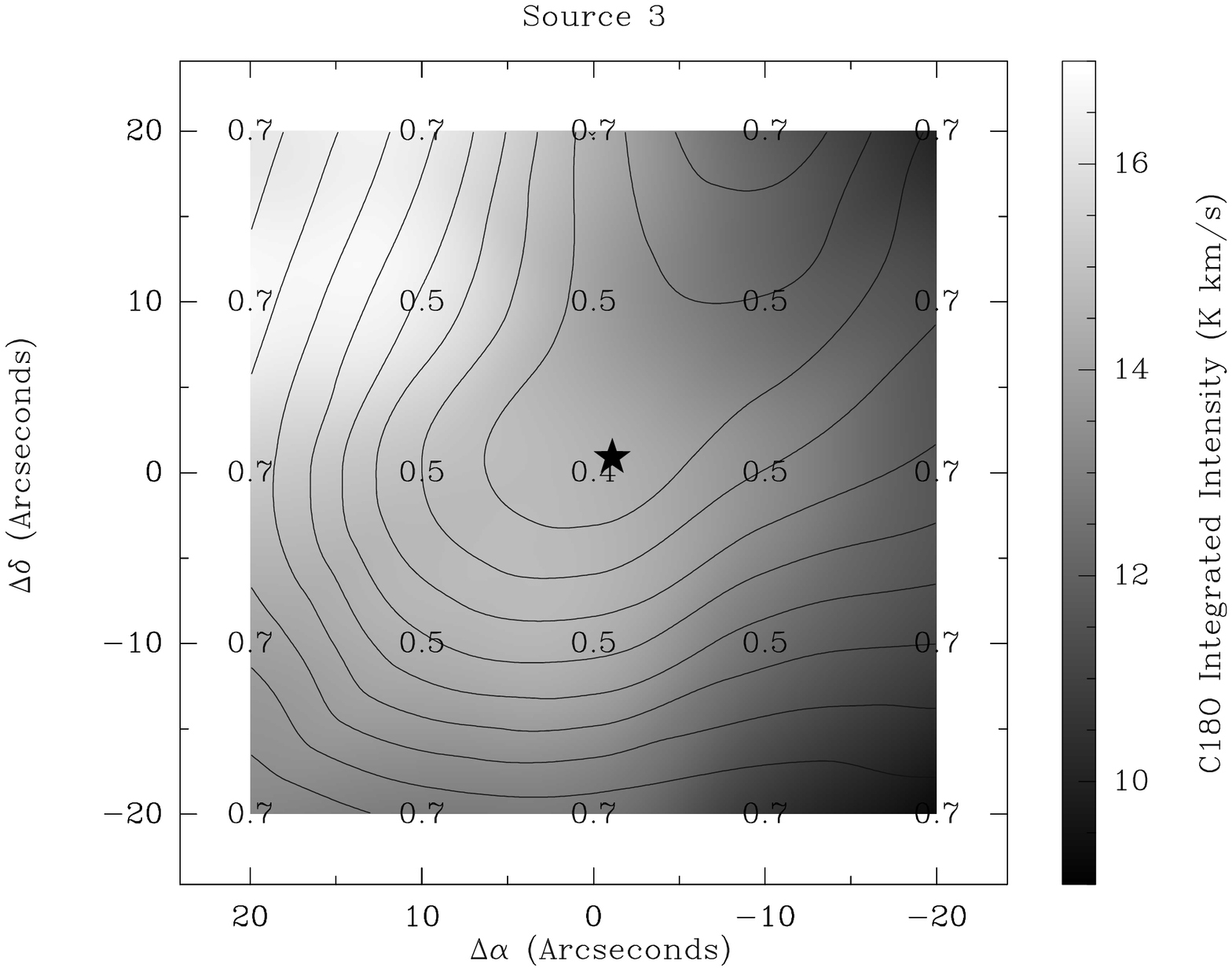} 
   \caption{Contour images of 850 $\mu$m continuum from SCUBA (in units of Jy beam$^{-1}$) smoothed to the CO beamsize and superimposed on C$^{18}$O integrated intensity maps (gray scale) in units of  $\int{T_A^*dV}$ for  Source 3. Contour levels are from 1.07 Jy beam$^{-1}$ to 5.47 Jy beam$^{-1}$ in steps of 0.4 Jy beam$^{-1}$. The beam HPBW of the C$^{18}$O observations is $\approx$ 22.8$''$. The numbers indicate the ratio of column density calculated from the C$^{18}$O to column density calculated from the dust. The star marks the position of the Spitzer source.} 
   \label{fig:Source3_con}
\end{figure}

\begin{figure}[h] 
   \centering
 \includegraphics[angle=270, width=5in]{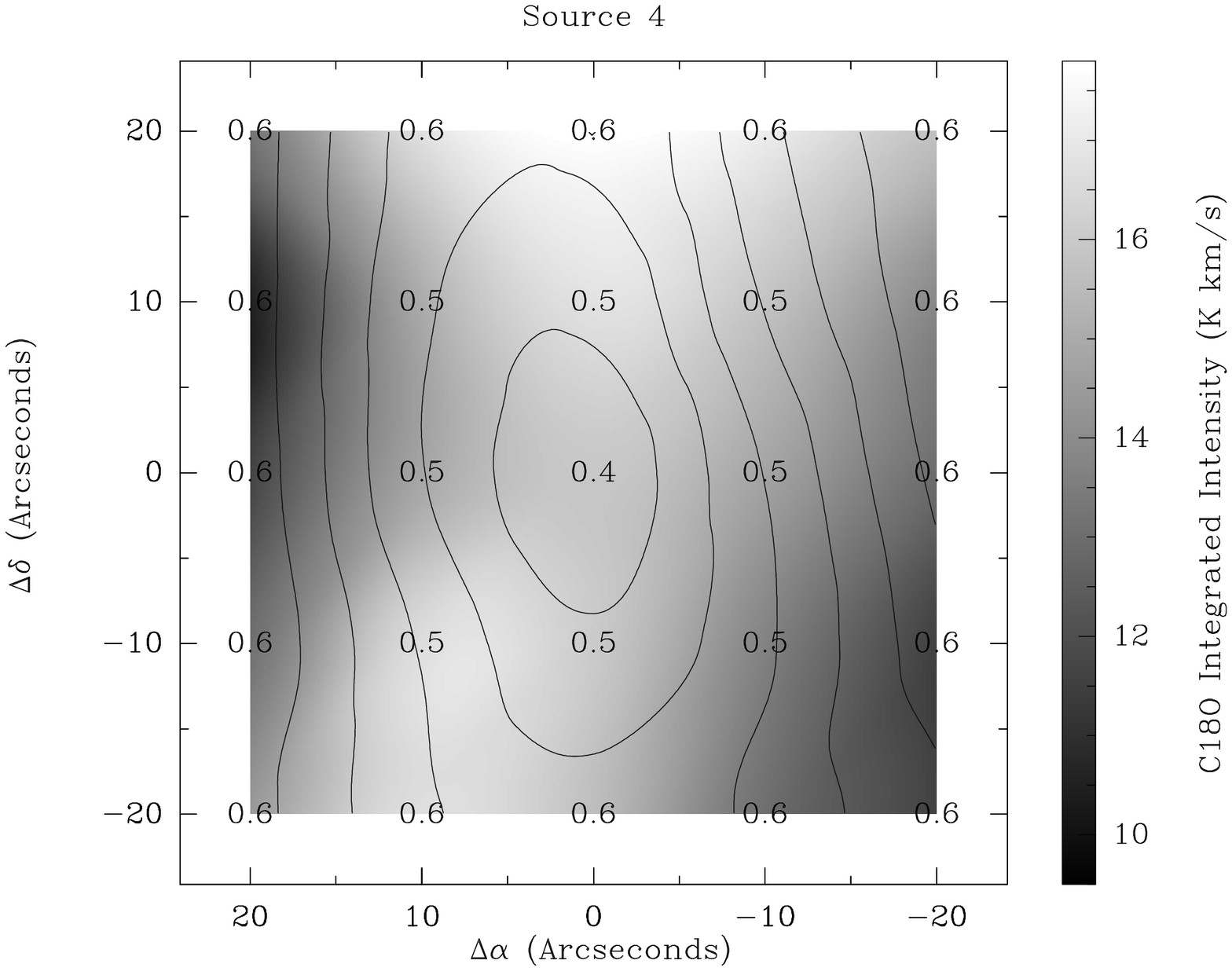} 
   \caption{Contour images of 850 $\mu$m continuum from SCUBA (in units of Jy beam$^{-1}$) smoothed to the CO beamsize and superimposed on C$^{18}$O integrated intensity maps (gray scale) in units of  $\int{T_A^*dV}$ for Source 4. Contour levels are from 1.94 Jy beam$^{-1}$ to 6.14 Jy beam$^{-1}$ in steps of 0.7 Jy beam$^{-1}$. The beam HPBW of the C$^{18}$O observations is $\approx$ 22.8$''$. The numbers indicate the ratio of column density calculated from the C$^{18}$O to column density calculated from the dust. } 
   \label{fig:Source4_con}
\end{figure}

\begin{figure}[h] 
   \centering
 \includegraphics[angle=270, width=5in]{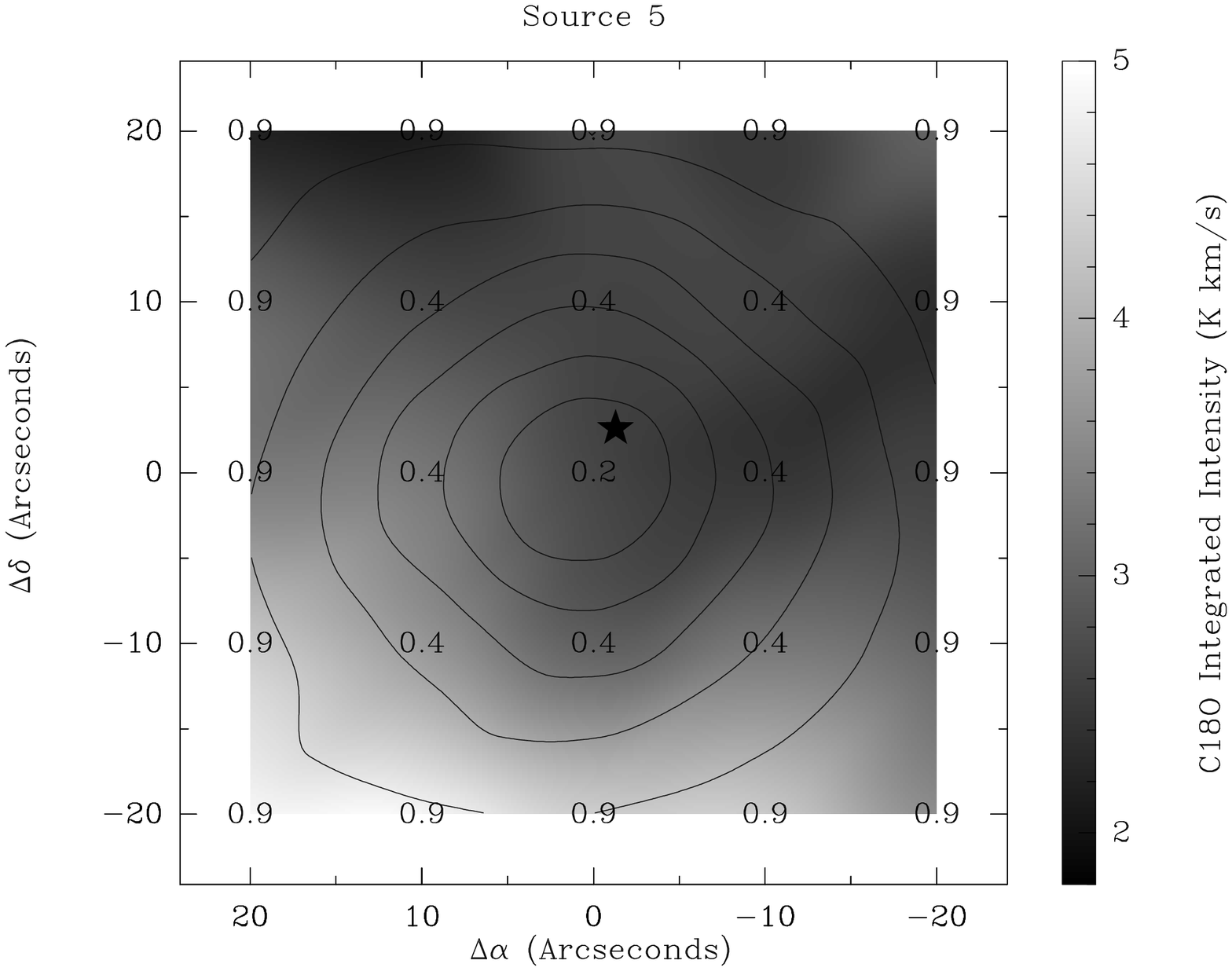} 
   \caption{Contour images of 850 $\mu$m continuum from SCUBA (in units of Jy beam$^{-1}$) smoothed to the CO beamsize and superimposed on C$^{18}$O integrated intensity maps (gray scale) in units of  $\int{T_A^*dV}$ for Source 5. Contour levels are from 0.1 Jy beam$^{-1}$ to 0.70 Jy beam$^{-1}$ in steps of 0.1 Jy beam$^{-1}$. The beam HPBW of the C$^{18}$O observations is $\approx$ 22.8$''$. The numbers indicate the ratio of column density calculated from the C$^{18}$O to column density calculated from the dust. The star marks the position of the Spitzer source.} 
     \label{fig:Source5_con}
\end{figure}

\begin{figure}[h] 
   \centering
 \includegraphics[angle=270, width=5in]{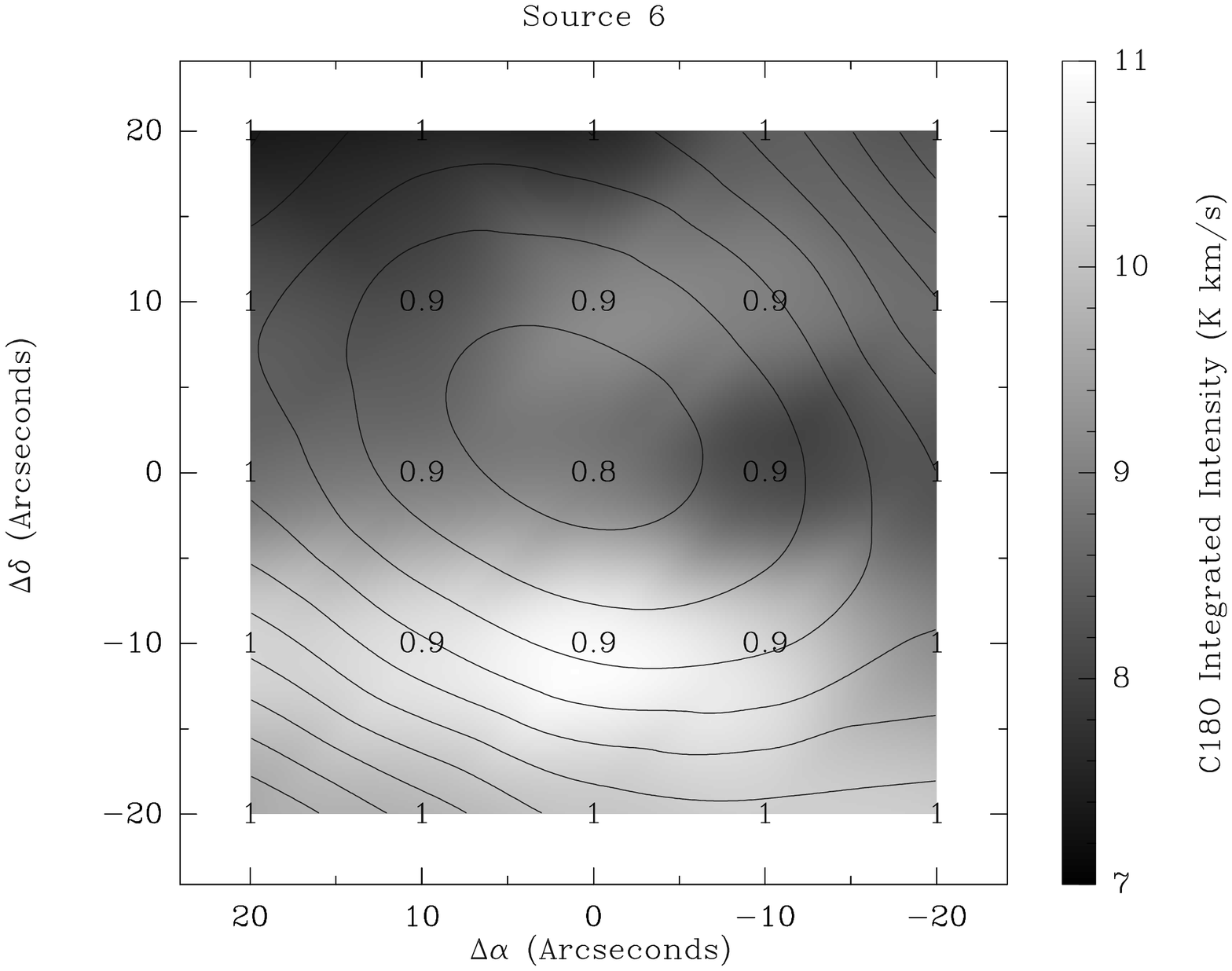} 
   \caption{Contour images of 850 $\mu$m continuum from SCUBA (in units of Jy beam$^{-1}$) smoothed to the CO beamsize and superimposed on C$^{18}$O integrated intensity maps (gray scale) in units of  $\int{T_A^*dV}$ for Source 6. Contour levels are from 0.64 Jy beam$^{-1}$ to 1.07 Jy beam$^{-1}$ in steps of 0.043 Jy beam$^{-1}$. The beam HPBW of the C$^{18}$O observations is $\approx$ 22.8$''$. The numbers indicate the ratio of column density calculated from the C$^{18}$O to column density calculated from the dust.}
   \label{fig:Source6_con}
\end{figure}

\begin{figure}[h] 
   \centering
 \includegraphics[angle=270, width=5in]{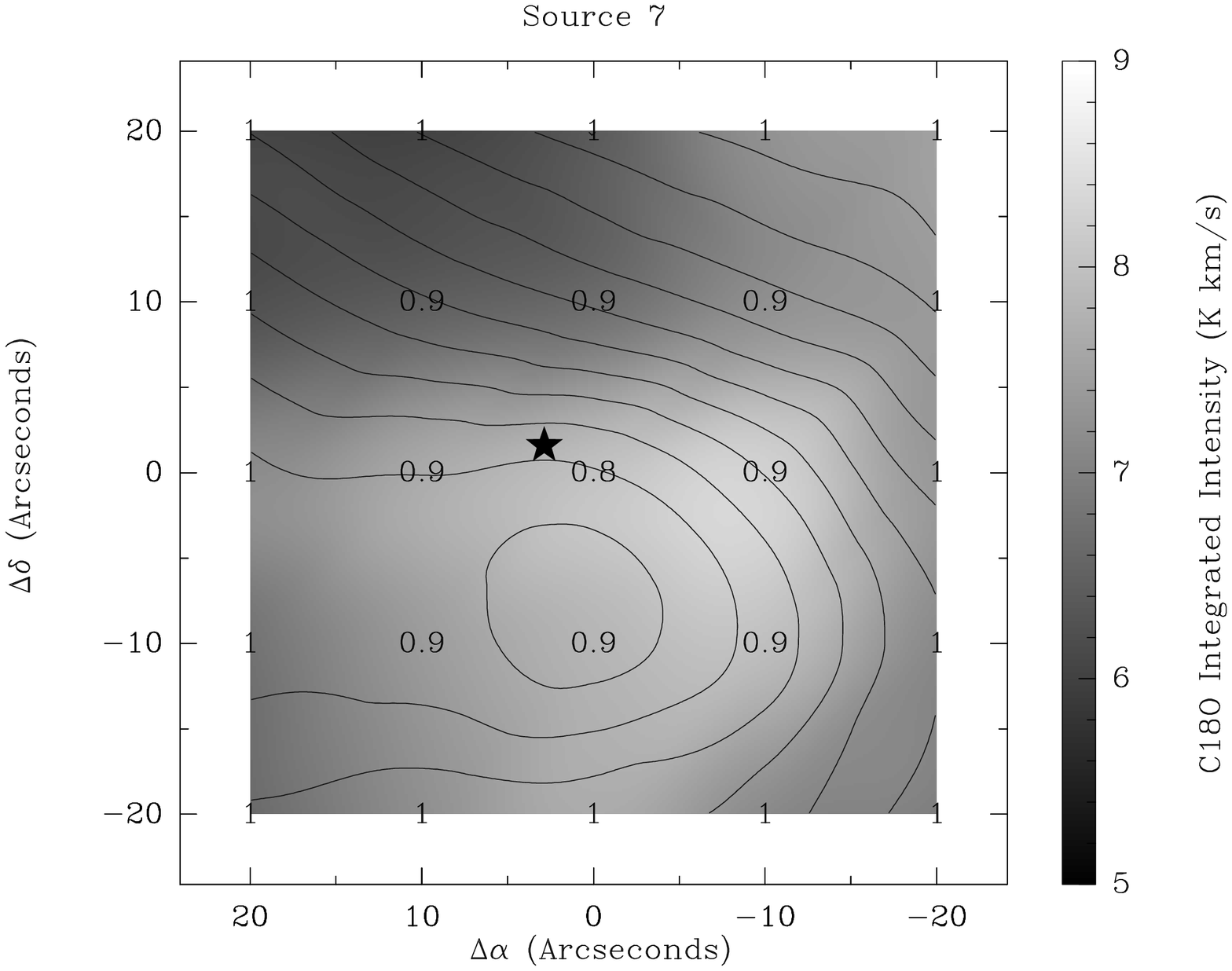} 
   \caption{Contour images of 850 $\mu$m continuum from SCUBA (in units of Jy beam$^{-1}$) smoothed to the CO beamsize and superimposed on C$^{18}$O integrated intensity maps (gray scale) in units of  $\int{T_A^*dV}$ for Source 7. Contour levels are from 0.24 Jy beam$^{-1}$ to 0.9 Jy beam$^{-1}$ in steps of 0.06 Jy beam$^{-1}$. The beam HPBW of the C$^{18}$O observations is $\approx$ 22.8$''$. The numbers indicate the ratio of column density calculated from the C$^{18}$O to column density calculated from the dust. The star marks the position of the Spitzer source.} 
   \label{fig:Source7_con}
\end{figure}

\begin{figure}[h] 
   \centering
 \includegraphics[angle=270, width=5in]{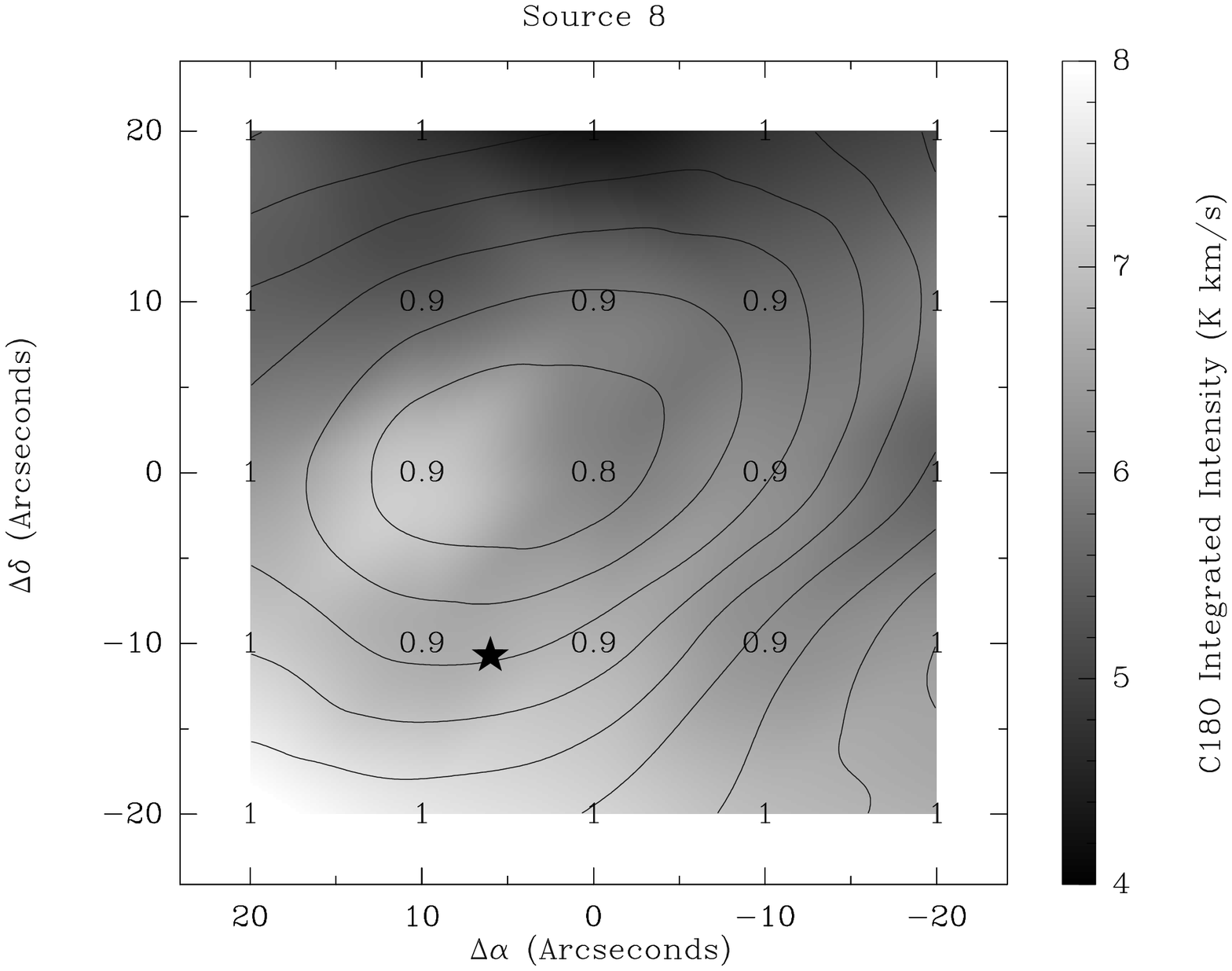} 
   \caption{Contour images of 850 $\mu$m continuum from SCUBA (in units of Jy beam$^{-1}$) smoothed to the CO beamsize and superimposed on C$^{18}$O integrated intensity maps (gray scale) in units of  $\int{T_A^*dV}$ for Source 8. Contour levels are from 0.55 Jy beam$^{-1}$ to 1.07 Jy beam$^{-1}$ in steps of 0.065 Jy beam$^{-1}$. The beam HPBW of the C$^{18}$O observations is $\approx$ 22.8$''$. The numbers indicate the ratio of column density calculated from the C$^{18}$O to column density calculated from the dust. The star marks the position of the Spitzer source.}
   \label{fig:Source8_con}
\end{figure}

\begin{figure}[h] 
   \centering
 \includegraphics[angle=270, width=5in]{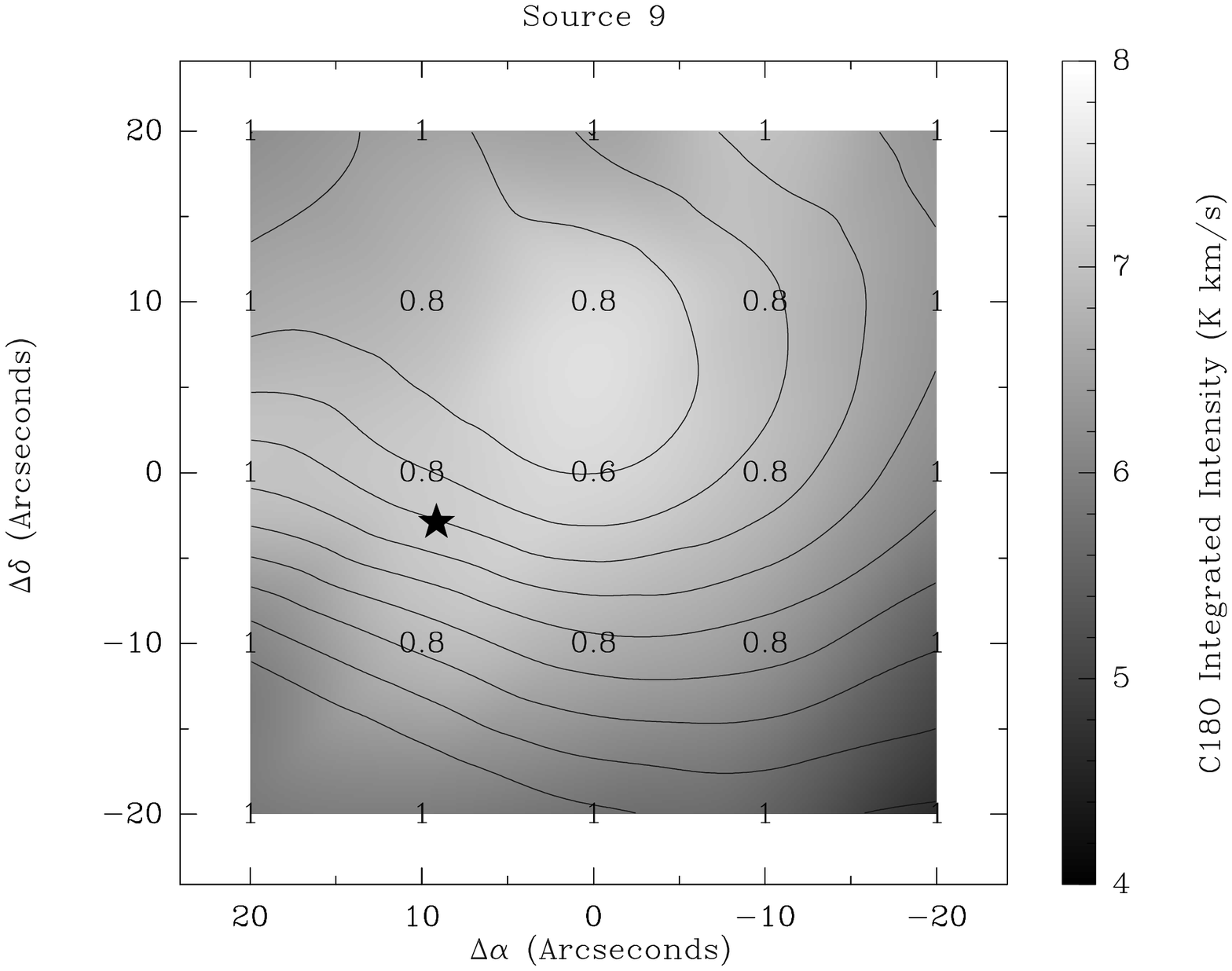} 
   \caption{Contour images of 850 $\mu$m continuum from SCUBA (in units of Jy beam$^{-1}$) smoothed to the CO beamsize and superimposed on C$^{18}$O integrated intensity maps (gray scale) in units of  $\int{T_A^*dV}$ for Source 9. Contour levels are from 0.01 Jy beam$^{-1}$ to 0.8 Jy beam$^{-1}$ in steps of 0.08 Jy beam$^{-1}$. The beam HPBW of the C$^{18}$O observations is $\approx$ 22.8$''$. The numbers indicate the ratio of column density calculated from the C$^{18}$O to column density calculated from the dust. The star marks the position of the Spitzer source.} 
   \label{fig:Source9_con}
\end{figure}

Figures \ref{fig:Source1_con} to \ref{fig:Source9_con} show, for each source, the SCUBA continuum maps smoothed to the CO beamsize (contours) 
superimposed on the C$^{18}$O integrated intensity maps (gray scale) in units of $\int{T_A^*dV}$. 
The SCUBA continuum maps show very clear intensity peaks at the map center, with the exception of 
Sources 3 and 9, which suffer from confusion with nearby more intense (but less compact) cores. In contrast, 
the C$^{18}$O emission is distributed quite smoothly across the maps. 

In the following three subsections we analyze the observational case for molecular outflows, gas infall, 
and CO depletion in this data set.

\subsection{Molecular Outflows}  
\label{sec:outflow}

 Molecular outflows are high velocity, collimated streams of ambient circumstellar material that has been 
swept up in the stellar wind from a young protostar. The driving mechanisms behind these outflows are 
believed to be highly collimated stellar jets of ionized and neutral atomic material (Bachiller 1996), 
with opening angles as small as 3 - 5$^{\circ}$, lengths of 10$^3$ - 10$^4$ AU, and velocities between 
150 - 400 km s$^{-1}$ (K\"{o}nigl \& Pudritz 2000; Arce et al. 2007; Andr\'{e} et al. 2008). Circumstellar molecular material is caught up in 
these jets and winds, either through entrainment along the sides of the jets, or through interface bow 
shocks in the jet (Klaassen et al. 2006; Bachiller 1996). The CO gas forms highly collimated streams along the flow axis at 
high velocities, and less collimated flows (collimation factors between 3 and 10) at low velocities $\approx$ 30 km s$^{-1}$ 
(K\"{o}nigl \& Pudritz 2000; Andr\'{e} et al. 2008; Richer et al. 2000). Outflows are most commonly bipolar, although monopolar (NGC 2264, Lada 1985) 
and multipolar outflows (Mizuno et al. 1990; Avery et al. 1990) have been observed. Many of the multipolar outflows are 
believed to be part of bipolar outflows from other nearby sources (Bachiller 1996; Carrasco-Gonz\'{a}lez et al. 2008).

All low-mass YSOs have outflows as long as they are surrounded by a circumstellar envelope 
(Bontemps et al. 1996). The youngest class of protostars, Class 0, experience the strongest and most highly 
collimated bipolar outflows. As they move onto the next stage, Class I, the outflows become less energetic 
and less collimated (Arce \& Sargent 2006; Tobin et al. 2007; Bontemps et al. 1996).  As the protostar evolves further and accretes or dissipates its 
circumstellar envelope, it no longer has molecular outflows associated with it. Outflows then can be used as 
 tracers of evolutionary stages. If a molecular outflow is present, the source is most likely Class 0 or I. If 
there is no molecular outflow but there is depletion (see below), then the source is probably prestellar.

\begin{figure}[h] 
   \centering
 \includegraphics[angle=270, width=5in]{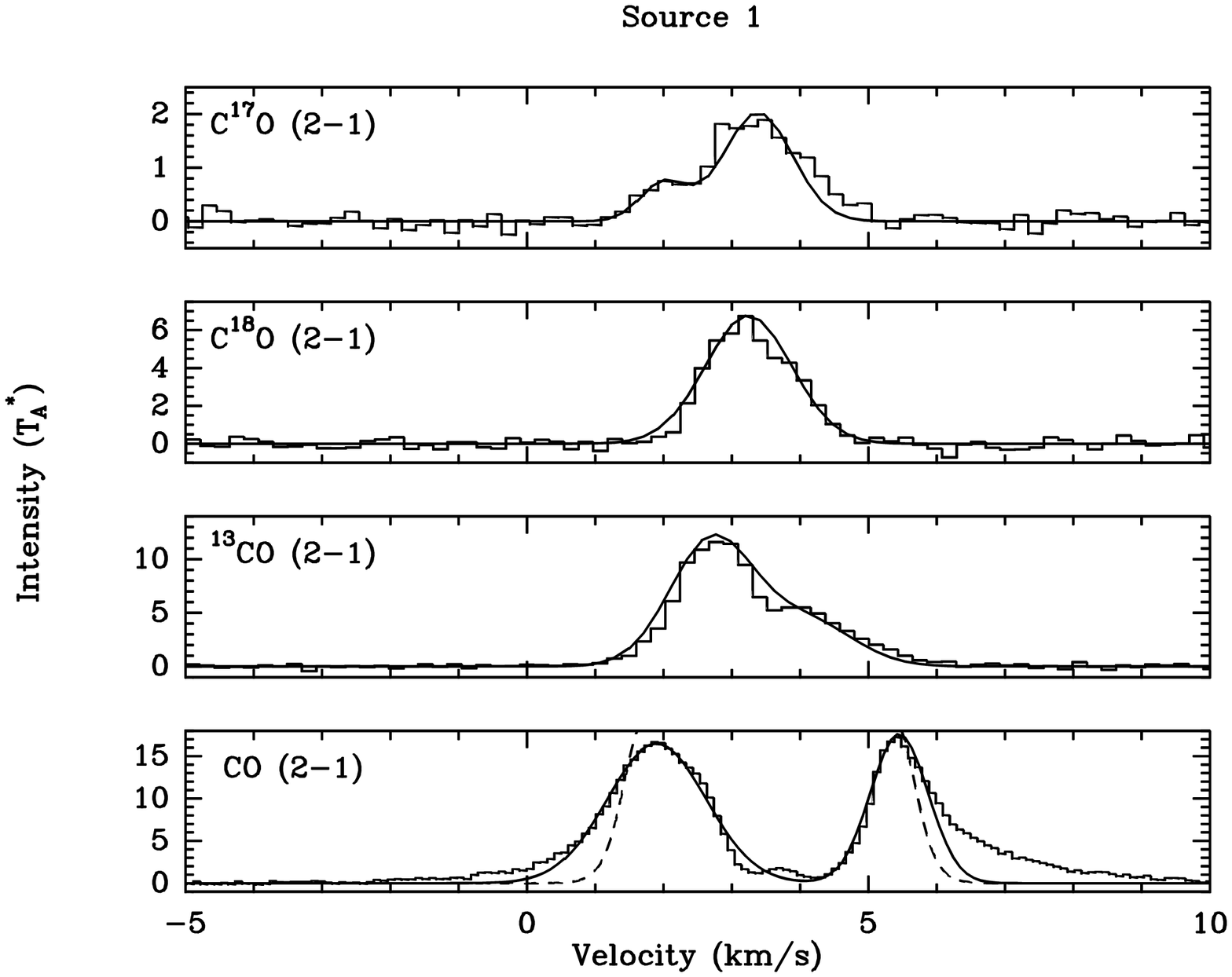} 
   \caption{From top to bottom: C$^{17}$O, C$^{18}$O, $^{13}$CO, and CO J = $2\rightarrow1$ emission from the central position in Source 1. The continuous curves in some of the panels represent single or multiple Gaussian profiles that were fit to the observed data, the results of which were used to determine the C$^{18}$O centroid velocity, integrated intensity, and column density, as well as the extent of the CO line wings (see Table \ref{tab:colden}). The dashed line in the bottom panel is an LTE model of an optically thick CO line.}
   \label{fig:Source1_spec}
\end{figure}

Because CO is the most abundant isotopologue  [i.e., X(CO)/X(H$_2$) $\approx 10^{-4}$], the relatively low 
column density in outflowing gas is often most easily traced through CO emission.   In addition, the 
physical conditions in molecular outflows are sufficient to excite CO J = 2$\rightarrow$1 emission
given its low excitation energy (16.6 K) and moderate critical density (n $\approx 10^4$ cm$^{-3}$).
Therefore, we primarily look for signs of molecular outflows in our cores using the CO observations. To quantitatively assign an outflow to a CO line, we fit Gaussians to the portions of the CO emission that were not self-absorbed. We found Gaussians fit in this way better fit the emission line wings than a single Gaussian fit to the entire CO spectrum. An example of such a single CO Gaussian fit is done for Source 1 in which we produced an LTE model of an optically thick CO line (Figure \ref{fig:Source1_spec}; bottom panel, dashed line) with a kinetic temperature of 20 K, a CO column density of 3.86 $\times 10^{18}$ cm$^{-2}$, and a linewidth of 1.5 km s$^{-1}$, which matches the linewidth of the optically thin tracer, C$^{18}$O. As can be seen in Figure \ref{fig:Source1_spec} the single component fit has a harder time fitting the emission in the line wings. Therefore, by using a two-component fit, we are producing the most conservative estimate of the outflow properties.  We consider an outflow present if the CO emission has an extension of at least 1 km s$^{-1}$ beyond where the CO Gaussian fit dips below the rms, and if the integrated intensity in the wing has a signal to noise ratio greater than 5. Gaussians were also fit to the C$^{18}$O at the central V$_{LSR}$ to measure the extent of the line wing in CO from the central velocity of the core. In cases where C$^{18}$O has multiple components, each component was fit and the V$_{LSR}$ was taken to be at the center of all the emission. Columns (2) and (4) of 
Table \ref{tab:colden} denote the full extent of the blue and red outflows, respectively, from the center of the optically thin C$^{18}$O to where the line wing emission in CO drops below the rms noise, while Columns (3) and (5) 
show the signal-to-noise ratio in the wings.

\subsection{Infall}
\label{sec:infall}
Radiative transfer models reveal that cores collapsing to form a star produce distinctive spectral profiles 
dependent on the optical depth of the emission line and the infall velocity.  Models show that in a collapsing core, optically thin lines produce a single Gaussian spectrum centered at the natal velocity of the core whereas optically thick lines produce double-peaked spectral line profiles with the blue peak brighter than the red peak. As the infall velocity increases, the degree of asymmetry increases as well, from two equally bright blue and red peaks to a brighter blue than red peak (Leung \& Brown 1977; Myers et al. 1996).  In principle,
therefore, we can use spectral line profiles to trace infall motion.

Mardones et al. (1997) quantify the asymmetry of the line profile using an infall parameter:
\begin{equation}
\delta V = (V_{thick} - V_{thin})/\Delta V_{thin},
\end{equation}
where V$_{thick}$ is the velocity of the more intense peak in the optically thick line, V$_{thin}$ is the 
velocity of the peak of the optically thin line, and $\Delta V_{thin}$ is the width (FWHM) of the optically 
thin line.  The normalization factor ($\Delta V_{thin}$) is introduced to account for the fact that cores 
have a range of line widths. Mardones et al. (1997) specify that an infall parameter of $\delta V < -0.25$ 
signifies a core that is undergoing significant infall.  
 
Prestellar cores can be defined as gravitationally bound objects within larger Giant Molecular Clouds that do not contain IR sources (Andr\'{e} et al. 2000). There is evidence that these cores begin the process of 
kinematic infall at this stage. Lee et al. (1999) conducted a survey in which they detected 69 
starless cores in the CS J = 2$\rightarrow$1 and N$_2$H$^+$ J = $1\rightarrow0$ lines. In a subset of 67 of these cores they found that 20 cores showed a blueshifted asymmetric 
profile, three showed red asymmetry, and the rest were symmetric. The mean $\delta$V for this sample of starless 
cores was -0.24 $\pm$ 0.04. Later it was found in a sample of 17 starless cores surveyed in the HCO$^{+}$
J = 3$\rightarrow$2 line that six cores had blue asymmetry, eight were symmetric, and three were not 
detected. A mean $\delta$V of -0.34$\pm$0.13 was found in this sample (Gregersen \& Evans 2000). Sohn et al. (2007) observed a sample of 64 starless cores in HCN J = 1$\rightarrow$0 hyperfine transitions. In a subset of 49 cores they found a mean $\delta$V of -0.51$\pm$0.06, -0.59$\pm$0.08, and -0.41$\pm$0.10 for the hyperfine components F = 0-1, F = 1-1, and F = 2-1 respectively. 
 
Infall has also been observed in Class 0 and Class I protostellar sources. Gregersen et al. (1997) surveyed 
23 Class 0 sources in the HCO$^{+}$ J = 4$\rightarrow$3 and J = 3$\rightarrow$2 lines. They found nine sources 
displayed blue asymmetry in their lines. Mardones et al. (1997) performed a CS J = 2$\rightarrow$1 and 
H$_2$CO J = $2_{12}\rightarrow1_{11}$ study of 23 Class 0 and 24 Class I objects. Fifteen of these sources were in common 
with Gregersen et al. (1997). They found that 25\% more sources had blue asymmetry than red, but almost all 
of the sources with blue asymmetry were Class 0 objects. In the H$_2$CO J = $2_{12}\rightarrow1_{11}$ line they found the mean $\delta$V in their sample to be 
-0.28$\pm$0.13 for Class 0 sources and $\delta$V = -0.02$\pm$0.07 for Class I sources. According to this study 
infall appears to end before the Class I stage. However, the trend may be due to varying line conditions within
the protostellar core. The CS and H$_2$CO lines are less opaque than the HCO$^{+}$ lines which trace the late stages 
of infall (Mardones et al 1997). Gregersen et al. (2000) attempt to address this issue by observing 18 Class 0 
sources and 16 Class I sources, finding the blue excess of both classes to be 0.31, that is displaying no difference 
in infall between the classes. Considering the mean $\delta$V, Gregersen \& Evans (2000) found for starless cores 
$\delta$V = -0.34$\pm$0.13, for Class 0 $\delta$V = -0.11$\pm$0.09, and for Class I $\delta$V = -0.17$\pm$0.08. 
Thus, on line asymmetry alone it appears difficult to distinguish between prestellar, Class 0 and Class I objects. 
The presence of infall does, however, imply that a core is undergoing gravitational collapse and likely will
form an embedded star.
 
For this study, to calculate $\delta$V we use $^{13}$CO as the optically thick line (CO itself is too self-absorbed), 
and C$^{18}$O as the optically thin line (e.g., Fig. \ref{fig:Source4_spec}). Column (6) of Table \ref{tab:colden} provides 
the infall parameter $\delta$V at the central core position only. Of the six cores for which we were able to calculate $\delta$V, all except Source 7 show evidence for significant infall. A dash in the infall column indicates that infall could not be calculated for 
this core due to confusion with multiple spectral features. However, although Source 7 and Source 8 have multiple spectral components, we are able to calculate infall parameters for these two cores since the multiple spectral components are separated enough to distinguish clearly the blue asymmetry of the core emission line. Column (7) list the central V$_{LSR}$ for each of the cores as taken from Gaussian fits to  the C$^{18}$O line.

\subsection{CO Depletion}
\label{sec:depletion}

At the low temperatures and high densities associated with prestellar and protostellar cores, CO will deplete 
from the gas phase by freezing onto dust grains.  J\o rgensen et al. (2005) suggests that the temperature below
which CO cannot evaporate back into the gas phase is  $\approx$ 35 K, and the density at which freezeout onto grains 
begins is $1 \times 10^4 - 6 \times 10^5$ cm$^{-3}$. 
  
At the densities and temperatures within typical prestellar objects, CO depletion is expected to occur within 
$\approx 10^5$ yr.  Thus, a transient object would not have sufficient time (or perhaps even density) for 
depletion to occur. Likewise, for protostellar cores, the characteristic depletion factor should correlate with the
protostellar age, since the depletion disappears during the latter Class I stage.

To estimate the amount of depletion we calculate the H$_2$ column density as derived from the molecular line 
observations and compare this value with  the H$_2$ column density determined from the submillimeter continuum 
observations. For the molecular line observations, column density is calculated from C$^{18}$O assuming the C$^{18}$O line to be optically thin, and in local 
thermodynamic equilibrium (LTE). This is a reasonable assumption given the low critical density of C$^{18}$O 
(2.4 $\times$ 10$^4$ cm$^{-3}$) and the low excitation temperature (15.8 K) required to excite the line. Using 
these assumptions, we calculate the column density of the upper state of the C$^{18}$O using the standard equations:

\begin{equation}
{
N_{{\rm C}^{18}{\rm O}} = \frac{N_2}{g_2} \frac{ \sum_{i=1}^\infty g_i e^{-E_i/kT}}{e^{-E_{21}/kT}},
}
\end{equation}\\
where

\begin{equation}
{
N_2 = \frac{8k\pi\nu^2}{hc^3} \frac{1}{A_{21}}\int T_R^* dV
}
\end{equation}
\\
is the column density in the upper state J=2, $\int T_R^* dV$ is the C$^{18}$O J = 2$\rightarrow$1 integrated 
intensity (corrected for the beam efficiency of 0.69), and T is the gas/dust temperature taken from J\o rgensen et al. (2008). The resultant H$_2$ column density is then calculated using a standard \hh/\ceo\ abundance ratio 
of  $5 \times 10^6$ (Wannier 1980).  \hh\ column densities calculated using this method are provided in Column (9) of Table \ref{tab:colden}.

The H$_2$ column density is also calculated from the dust continuum flux by using the usual equation: 

\begin{equation}
{
N_{{\rm H}_2} = \frac{S_\nu / \Omega_{beam}}{\mu m_H \kappa_{\lambda} B_{\nu}(T)}
}
\end{equation}\\
where $S_\nu$ is the flux from the 850 $\mu$m SCUBA observations, T is again taken from J\o rgensen et al. (2008), and

\begin{equation}
{
\kappa_{\lambda} = \kappa_0 (\frac{\lambda}{\lambda_0})^{-\beta}.
}
\end{equation}
\\
The value of $\kappa_0$ though somewhat uncertain has been estimated by various authors to be  
$\kappa_0 = 0.01\,$cm$^2\,$g$^{-1}$ at  $\lambda_0 = 1.3\,$mm and $\beta = 1.5$. This value
is in agreement with models of dust opacity in protostellar envelopes in the Ophiuchus A core. 
(Andr\'{e} \& Montmerle 1994; Andr\'{e} et al. 1993). Column densities calculated by this method are given in 
Column (10) of Table \ref{tab:colden}. 

Figures \ref{fig:Source1_con} to \ref{fig:Source9_con} also show the ratio of the average H$_2$ column density calculated from the gas to the average H$_2$ column density calculated from the dust (the numbers at each observed position provide the average column density ratio at that position). To perform this analysis, first, each core was split into three separate regions, the outer 16 points on the figures representing one annulus, the inner 8 points representing a second annulus, and then the single central point.  The ratio in each region/annulus was then calculated by taking the average column density as calculated from the C$^{18}$O emission within that annulus and dividing by the average column density as calculated from the 850 $\mu$m flux within the same annulus.  This was then repeated for the inner annulus of the core and at the center of the core. In six cases (Sources 1, 2, 6, 7, 8, and 9) this ratio was greater than unity in the outer annulus due to the subtraction of large scale structure in the SCUBA maps. In these cases a constant amount of continuum flux was added across the entire core to make the ratio unity at the outer annulus. A ratio of less than one at the peak in the SCUBA continuum flux indicates depletion at the core. All cores show some signs of depletion.

In order to quantify the degree of depletion, we define the depletion factor as the H$_2$ column density ratio at the edge divided by the H$_2$ column density ratio at the center of the core. A number of authors (Bacmann et al. 2002; Caselli et al. 1999; Willacy et al. 1998; Crapsi et al. 2004) have found CO depletion factors of 4 - 15 in samples of prestellar cores. However, because our cores may include (and do include) protostellar cores and, because we added continuum flux across the entire core in several cases rather than subtracting CO emission (which means that the depletion we measure is through the entire cloud rather than within only the core), we arbitrarily select a lower depletion factor (i.e., a depletion factor of $>$ 2) to indicate significant depletion. The depletion factors for our nine cores are discussed in detail in
\S 4.1. Hence only Source 2 and Source 5 appear to have significant depletion, while the rest of the cores have very little.

\section{Discussion}

Through comparison of the evidence for outflow, infall, and CO depletion, we can attempt to deduce the
evolutionary state of each core in our sample.  Sources are placed into three broad evolutionary classes: (a) inactive, in which the cloud shows no evidence for any star-formation activity, although molecular depletion may or may not occur depending on the age and physical conditions in the core.  (b) Pre-protostellar, in which the cloud shows evidence for infall and depletion but has no outflow activity since a central object has not yet formed.  (c) Protostellar, in which we believe a protostellar object is in the process of forming as indicated by the presence of an outflow.  In this case, infall or depletion may be present depending on the age of the protostar.  We would expect the youngest cores to display both, because the core must be collapsing to form the central object and may not have had enough time to heat up and evaporate the ice mantles (J\o rgensen et al. 2005).  The eldest cores, on the other hand, would not show evidence for either, because cloud collapse onto the mature protostar would have been halted and the gas temperatures will be high enough to evaporate the ice mantles and drive hot core/corino chemistry (J\o rgensen et al. 2005).  The metric for classifying sources as inactive, pre-protostellar, or protostellar is provided in Table \ref{tab:metric}.  

To test how well CO works as a tracer of star-formation activity, we compare our classification of each core based on the spectra of the CO isotopologues to the classification based on a comparison of Spitzer (MIPS and IRAC) and 850 $\mu$m SCUBA continuum observations. For the latter classification, a core is considered to be protostellar if the IRAC and MIPS data exhibit the colors [3.6] - [4.5] $>$ 1 and [8.0] - [24] $>$ 4.5, in which [3.6], [4.5], [8.0], and [24] are the magnitudes of the detected Spitzer sources at 3.6, 4.5, 8.0, and 24 $\mu$m, respectively. In addition, the cores must have concentrations of C $>$ 0.6, and the Spitzer MIPS source must be less than $15''$ from the nearest SCUBA core. This follows the treatment of J\o rgensen et al. (2007) for the Perseus complex. The Spitzer positions are taken from J\o rgensen et al. (2008). Column (6) of Table \ref{tab:class}, as well as the stars on individual Figures \ref{fig:Source1_con} to \ref{fig:Source9_con}, 
show the distance between the observed Spitzer protostars and the center of each SCUBA core. Results are presented in detail below and summarized in Table \ref{tab:class}. 

\subsection{Evidence for Star Formation Activity}
\label{sec:sf}

\textbf{Source 1}\\
Source 1 is identified by Spitzer as containing a protostar 10.2$''$ from the submillimeter peak.  This object is associated with the YSO identifiers J162622-242254, GSS-30 IRS 2 and VSSG 12 (Barsony et al. 1997; Grasdalen et al. 1973; Vrba et al. 1975) and is classified as a Class III object (Allen et al. 2002). However Allen et al. 2002 suggest that the evolved state of this object indicates that it is most likely unrelated and foreground to the cloud. Hence this core may in fact be starless. From the C$^{18}$O observations we found that Source 1 has only one velocity component at about 3.2 km s$^{-1}$. There is evidence for a small amount of depletion in the core (Fig. \ref{fig:Source1_con}); the depletion factor (defined in \S 3.3) is 1.52 (Table \ref{tab:class}). Figure \ref{fig:Source1_spec} also shows definite line wings present in the CO at the center position of the core, which indicates both strong blue and red outflows. The blue line wing in CO extends over about 5.6 km s$^{-1}$ and the red line wing over 8.1 km s$^{-1}$ from the center velocity of the C$^{18}$O spectra.  In addition, the infall parameter for this core is -0.361 suggesting significant infall motions in this core. Thus we conclude that this core should contain a protostar, however we are unable to confirm this with Spitzer observations. 

\textbf{Source 2}\\
Spitzer identified this core as containing the T-Tauri star J162624-241616 (Barsony et al. 1997) which is identified to be a Class II object (Bontemps et al. 2001).  Our C$^{18}$O observations show evidence for two separate velocity components - one at  2.5 km s$^{-1}$ and the other at 3.4 km s$^{-1}$. The \hh\ ratio map (Fig. \ref{fig:Source2_con}) shows evidence for significant depletion in the core, with a depletion factor of 3.33. In addition, we see evidence of both strong blue and red outflows as shown in Figure \ref{fig:Source2_spec} with a blue line wing that extends over 6.0 km s$^{-1}$ and a red line wing that extends over 6.5 km s$^{-1}$. The infall parameter for this core could not be calculated due to confusion with the secondary C$^{18}$O spectral feature. However, the high depletion and strong outflows indicate that this core contains an embedded protostar, which is consistent with the Spitzer observations.
 
\textbf{Source 3}\\
Spitzer observations reveal the presence of the well-known Class 0 object VLA 1623 (Andr\'{e} et al. 1990). We observe only one C$^{18}$O spectral feature in this core, at 3.7 km s$^{-1}$.  The blue line wing in the CO spectra extends over 7.4 km s$^{-1}$ and the red line wing over 6.9 km s$^{-1}$, indicating a strong blue outflow and a weaker red outflow, as seen in Figure \ref{fig:Source3_spec}. The \hh\ ratio map (Fig. \ref{fig:Source3_con}) shows evidence for depletion at the center of the core, with a depletion factor of 1.75. There are also strong infall motions in this core, with an infall parameter of -0.696 at the center. The strong infall and outflows would indicate that this core does indeed contain a protostar, again in agreement with the Spitzer observations. 

\textbf{Source 4}\\
This core is starless because no Spitzer source was detected within 15$''$.  However, the molecular line observations would indicate that this core contains a protostar from the following observations. 
The \hh\ ratio map (Fig. \ref{fig:Source4_con}) shows evidence for depletion at the center of the core, although it is small with a depletion factor of only 1.50. There is also evidence for strong infall in this core with an infall parameter of -0.682.  Figure \ref{fig:Source4_spec} shows a red line wing that extends a little over 4.4 km s$^{-1}$, indicating the presence of a red outflow. There is no blue line wing seen in this core.  Thus, while we would classify this object as protostellar, this is at odds with the Spitzer results. However, as this core is very near Source 3, it is possible that the outflow from Source 3 is impacting the region and the outflow we are seeing is actually from Source 3. If this is the case then Source 4 has no outflow but does have infall and weak depletion classifying it as pre-protostellar, which is in agreement with Spitzer results.

\textbf{Source 5}\\
A protostellar object is identified in Source 5 by Spitzer. The molecular line observations show spectral features at 3.3 km s$^{-1}$, which we observe in all molecular lines, and one at 4.4 km s$^{-1}$, which we see in the $^{13}$CO line (Fig. \ref{fig:Source5_spec}). A red outflow extending approximately 3.2  km s$^{-1}$ is observed in this core, as well as a weak blue outflow. There is also significant depletion in this core as shown in the \hh\ ratio map (Fig. \ref{fig:Source5_con}) with a depletion factor of 4.50. The infall parameter for this core is -0.526, indicating there is significant infall in this core. Taking into account the outflow, strong infall, and strong depletion, we classify this core as containing an embedded protostar. Spitzer confirms this identification and, in fact, Source 5 is a Class II YSO identified as J162645-242309, GSS 39, and VSSG 28 in a newly formed cluster of young stars (Barsony et al. 1997; Grasdalen et al. 1973; Vrba et al. 1975; Bontemps et al. 2001).

\textbf{Source 6}\\
Source 6 is starless according to Spitzer. In the molecular line observations we found that Source 6 displayed three spectral features at 2.5 km s$^{-1}$, 3.8 km s$^{-1}$, and 4.9 km s$^{-1}$, which we observe in all lines. Due to confusion with these multiple features, the infall parameter could not be calculated. There is very little evidence for depletion as the \hh\ ratio map (Fig. \ref{fig:Source6_con}) shows only a depletion factor of about 1.25. In addition, Figure \ref{fig:Source6_spec} shows no blue line wing present, but an extended red feature (extending for a little over 2.7 km s$^{-1}$) which may or may not be indicative of outflow.  With this inconclusive evidence we could not confidently classify this core using only molecular line tracers, but suggest it might be inactive. 

\textbf{Source 7}\\
Spitzer shows that Source 7 contains an embedded protostar identified as the Class I object  J162726-244051, rho Oph E5, or YLW15 (Barsony et al. 1997; Bontemps et al. 2001; Alexander et al. 2003; Young et al. 1986). However, in the molecular lines, this core does not display signs of being protostellar in nature. 
The depletion for this core as displayed in the \hh\ ratio map (Fig. \ref{fig:Source7_con}) is very small, with a depletion factor of 1.25. In addition, the infall parameter for this core is -0.213 suggesting little or no significant infall activity. The CO at the center position of the core shows no blue, and only very weak red line wings (that fall below our threshold: Table \ref{tab:colden} and Fig. \ref{fig:Source7_spec}). From all of these signs, we conclude that this core is likely inactive and starless, a conclusion which is at odds with the Spitzer results. 

\textbf{Source 8}\\
The Source 8 core contains the Class II object J162728-242721 and VSSG 18 (Bontemps et al. 2001; Barsony et al. 1997; Vrba et al. 1975) which is located 12.5$''$ from the submillimeter peak. 
There is evidence for only very weak depletion in this core (Fig. \ref{fig:Source8_con}), with a depletion factor of only 1.25. However, there are strong infall motions in this core as indicated by the infall parameter of -0.870. There is also a strong blue outflow (although no red outflow) as shown by the line wing in Figure \ref{fig:Source8_spec} that extends for more than 9.2 km s$^{-1}$. Although depletion is weak, this strong infall and outflow also leads us to conclude that this core contains a protostar.  

\textbf{Source 9}\\
Spitzer shows that Source 9 contains the Class II object J162730-242744 or Elias 33 (Bontemps et al. 2001; Barsony et al. 1997; Elias 1978) located 9.7$''$ from the submillimeter peak. The spectra for Source 9 (Fig. \ref{fig:Source9_spec}) show three separate velocity components at 2.6 km s$^{-1}$, 3.3 km s$^{-1}$, and 4.4 km s$^{-1}$, which are present in each molecular line.  We could not calculate infall for this core due to confusion between these three spectral features. The core shows blue and red line wings extending for about 5.5 km s$^{-1}$ and 3.6 km s$^{-1}$, respectively.  Figure \ref{fig:Source9_con} shows evidence for weak depletion in the core, with a depletion factor of 1.67. However, with the blue and red outflows and depletion, we classify Source 9 as being protostellar, consistent with the Spitzer results.

\subsection{Effectiveness of Molecular Line Tracers as a Probe of Star Formation}
\label{sec:spitzer}

Table \ref{tab:class} shows our classification of the nine cores based upon evidence for outflow, depletion, and infall,  versus the designation provided by the Spitzer data which identifies the closest IR source to the submillimeter core.  

Based on molecular line data alone, we classify 6 to 7 of the cores as protostellar (Source 4 having an uncertain classification) and one or two as being inactive (due to an uncertain classification for Source 6).  Three of our designations are clearly at odds with the Spitzer results: Source 4 we identified as being protostellar due to depletion, strong red outflow, and strong infall, whereas the closest IR source is further than 15$''$ from the submillimeter peak. Source 7 we identified as being inactive due to its lack of an outflow, weak infall, and weak depletion; however, Spitzer finds an IR source only 3.8$''$ from the submillimeter peak.  In addition, we identified Source 1 as being protostellar; however we are unable to confirm this result with Spitzer because the identified object may, in fact, be an evolved, foreground star. Thus, at first glance, it appears that using molecular line observations alone to classify submillimeter cores as protostellar or starless works about 67\% (6/9) of the time. However, if the outflow identified in Source 4 originates from Source 3, then we would classify Source 4 as pre-protostellar in agreement with Spitzer observations. This would make molecular line observations correct 78\% (7/9) of the time.

However, care must be taken before making such a naive statement.  First, we are obviously limited by small-number statistics and a larger sample would be needed to test this definitively.  Second, almost everywhere we have looked in Ophiuchus we have seen evidence for outflows and infall.  Without extensive mapping it is impossible to say whether, in a crowded field like that in Ophiuchus, the outflows are originating from an embedded YSO or from a YSO in an adjacent core.
Large mapping surveys like the GBS may be able to overcome this confusion and disentangle one outflow from another by mapping the full extent of the outflows and being able to find the originating object.  Our CO maps were limited in scope ($20'' \times 20''$) and so were unable to do this. In addition, the higher-energy CO J = $3\rightarrow2$ transition has a higher critical density and so will trace the warm dense gas in the outflow rather than the lower density surrounding cloud material. The higher resolution of the GBS observations at 345 GHz ($\theta_{FWHM} \approx 14''$ vs. our $22''$) may also provide a clearer picture of outflow activity in crowded fields.

 In addition, because the central V$_{LSR}$'s of the optically thin molecular species used to measure the infall parameter are so similar (Table \ref{tab:colden}) to each other, and to the native V$_{LSR}$ of the Ophiuchus cloud, our infall parameter may be measuring global motion of the cloud, rather than collapse of each individual core.  Thus, the infall parameter may not be an accurate tracer of star-formation activity.  This is especially true because we are using a ``nonstandard'' infall tracer - the low J transitions of CO and its isotopologues.  The low critical density and excitation temperature of of these transitions may mean that we are preferentially tracing the motions of the core surfaces, rather than their interiors, which are better probed by  transitions with higher critical densities.
 Even depletion is not a good measure of protostellar versus starless cores, because some of our largest depletion factors are seen in cores with confirmed IR objects.

Therefore, in conclusion, it seems that while CO observations alone can be (and have been) used fairly successfully to search for star-formation activity in isolated cores, great care must be taken in interpreting these results in crowded fields.  Multiple outflows from adjacent YSOs, global infall of the cloud, and a lack of correlation between depletion factors and embedded IR objects make it difficult to classify the cores with any degree of certainty.  While it is clear that useful information exists in these tracers, more detailed, higher-resolution observations are needed over moderately large spatial scales.  However, despite these caveats, it is remarkable that these simple and easily observable molecular lines seem to do a reasonably good job in  describing the crude evolutionary state of cores in star-forming regions.

\newpage

\begin{sidewaystable}[h] 
\caption{Source List}
\begin{center}
\begin{tabular}{lcccccc}
\hline\hline
Source & Designation & Johnstone et al. (2000)  & RA(J2000) & DEC(J2000) & Concentration & 850$\mu$m SCUBA peak flux $^{(b)}$ \\
    & & Designation  & (hh:mm:ss.s) & ($\circ$:$'$:$''$) & C$^{(a)}$ & (Jy beam$^{-1}$) \\
(1) & (2) & (3) & (4) & (5) & (6) & (7)\\
\hline 
 
Source 1 	        &	162622-24225	        &	         16263-2422              &        16:26:21.6	        &	-24:22:54.8	      &       0.58     &      0.82  	\\
Source 2 	        &	162624-24162	        &	 	16264-2416	        &        16:26:24.1	        &	-24:16:15.9	       &      0.41     &      0.45  \\
Source 3 	        &	162626-24243	        &		16264-2424	        &        16:26:26.5	        &	-24:24:30.9	       &      0.80     &      3.23   	\\
Source 4 	        &	162628-24235	        &		16264-2423	        &	16:26:27.5	        &	-24:23:54.9	       &      0.80     &      4.22    	\\
Source 5 	        &	162645-24231	        &	         16267-2423	        &	16:26:45.1	        &	-24:23:10.2	       &       0.74     &      0.67    	\\
Source 6 	        &	162660-24343	        &		16269-2434	        &        16:26:59.6	        &	-24:34:25.3	       &       0.55     &      0.59	\\
Source 7 	        &	162727-24405	        &		16274-2440	        &        16:27:26.7	        &	-24:40:52.3	       &       0.57     &      0.57   	\\
Source 8 	        &	162728-24271	        &		16274-2427b	        &        16:27:28.0	        &	-24:27:10.3	       &       0.34     &      0.66  	\\
Source 9 	        &	162729-24274	        &		16274-2427c	        &        16:27:29.5	        &	-24:27:40.3	       &       0.50     &      0.47   \\

 \hline \hline
\end{tabular}
\end{center}
(a) The concentration parameter C is defined in eq. (1) and described in \S 1.\\
(b) Peak flux calculated for the 14$''$ beam of the JCMT at 850 $\mu$m.
\label{tab:properties}
\end{sidewaystable}

\newpage

\begin{sidewaystable}[h]
\caption{Core Properties at the Central Position}
\begin{center}
\begin{tabular}{lccccccccccc}
\hline\hline
Source & Blue Outflow & Blue Outflow& Red Outflow & Red Outflow & $\delta$V & V$_{LSR}$ & N$_{C_{18}O}$ & N$_{H_2}$$^a$ & N$_{H_2}$$^b$\\
& Velocity Extent & S/N & Velocity Extent & S/N & & (C$^{18}$O)& &\\
&  (km s$^{-1}$) & &  (km s$^{-1}$) & & & (km s$^{-1}$)& (cm$^{-2}$) & (cm$^{-2}$) & (cm$^{-2}$)\\

(1) & (2) & (3) & (4) & (5) & (6) & (7) & (8) & (9) & (10)\\
\hline 
Source 1 	&       5.6	  &     10.47     &	 8.1         &       14.8        &       -0.361     &    3.24      &      $7.73\times10^{15}$             &         $3.86\times10^{22}$         	&	$5.86\times10^{22}$\\
Source 2 	&       6.0	  &     6.27     &	6.5         &       5.06        &         -               &    3.35      &      $3.38\times10^{15}$            &         $1.69\times10^{22}$    	&	$6.21\times10^{22}$\\
Source 3 	&	7.4	  &     16.45       &  6.9     &       21.55      &          -0.696    &    3.70      &      $1.18\times10^{16}$	    &         $5.88\times10^{22}$    	 &	$1.37\times10^{23}$\\
Source 4 	&	-	  &     -      &	4.4	     &       28.92      &         -0.682     &    3.65      &      $1.30\times10^{16}$	    &         $6.52\times10^{22}$    	&	$1.66\times10^{23}$\\
Source 5	&	3.2	  &     6.07 &	3.7    &        6.38              &             -0.526     &    3.42     &      $2.29\times10^{15}$            &         $1.15\times10^{22}$    	&	$8.75\times10^{22}$\\
Source 6	&	-	  &     -      &	2.7	     &        24.82      &        -               &     3.80     &      $6.95\times10^{15}$           &         $3.47\times10^{22}$    	&	$4.30\times10^{22}$\\
Source 7	&	-	  &    -      &	 4.5        &       3.69         &             -0.213     &     3.88     &      $6.25\times10^{15}$	    &         $3.12\times10^{22}$    	&	$3.75\times10^{22}$\\
Source 8	&	9.2	  &     16.54      &	-   & 	     -               &        -0.870     &      4.18     &      $5.31\times10^{15}$            &         $2.65\times10^{22}$        &	$3.33\times10^{22}$\\
Source 9	&	5.5     &     10.71     &	3.6          &      6.36         &         -              &      3.30     &      $5.93\times10^{15}$           &         $2.96\times10^{22}$	         &	$4.65\times10^{22}$\\
 \hline \hline
\end{tabular}
\end{center}

(a) Column densities calculated from $C^{18}O$ intensity.

(b) Column densities calculated from SCUBA flux.

\label{tab:colden}
\end{sidewaystable}

\newpage

\begin{table}[h]
\caption{Metric for Classification of Cores}
\begin{center}
\begin{tabular}{lccccc}
\hline\hline
Classification &  Outflow &  Infall  & Depletion \\
\hline 
Inactive	&	x  &  x  &  ?	\\
Pre-Protostellar & x & $\checkmark$  & $\checkmark$  \\
Protostellar	& $\checkmark$	&  ?  & ? \\
 \hline \hline
\end{tabular}
\end{center}
\label{tab:metric}
\end{table}

\newpage

\begin{table}[h]
\caption{Comparison of our Classification with Spitzer Data}
\begin{center}
\begin{tabular}{lccccc}
\hline\hline
Source & Our Classification  &  Outflow &  Infall  & Depletion Factor & Closest Spitzer Source\\
& & & & &  ($''$)\\
(1) & (2) & (3) & (4) & (5) & (6)\\
\hline 
 Source 1	&	protostellar	 &         Blue/Red        &    Yes    &       1.52                   &	10.2~$^a$	\\
Source 2	&	protostellar 	 &         Blue/Red         &      -     &         3.33                   &	2.5        \\
Source 3	&	protostellar	 &         Blue/Red         &     Yes   &       1.75                   &	1.0	\\
Source 4	&	protostellar/pre-protostellar	 &         Red                  &      Yes   &       1.50                   &	 -	\\
Source 5	&	protostellar         &         Blue/Red	&       Yes   &        4.50                      &	3.1	\\
Source 6	&	inactive/protostellar            &          Maybe Red   &       -     &        1.25                   &	-	\\
Source 7	&	inactive               &          No outflow	&      No    &        1.25                    &	3.8	\\
Source 8	&	protostellar	 &          Blue                &      Yes   &        1.25                   &	12.5	\\
Source 9	&	protostellar         &         Blue/Red        &       -    &        1.67                   &	 9.7	\\
 \hline \hline
\end{tabular}
 $^a$ May be a foreground object.
\end{center}
\label{tab:class}
\end{table}

\clearpage
\newpage

\clearpage

\newpage

\newpage

\newpage

\newpage

\newpage

\newpage

\newpage

\newpage

\newpage

\newpage

\newpage

\newpage

\newpage

\newpage

\newpage

\newpage

\newpage

\newpage

\newpage
\begin{thebibliography}{}

\bibitem[]{} Alexander, R. D., Casali, M. M., Andr\'{e}, P., Persi, P., \& Eiroa, C. 2003, A\&A, 401, 613
\bibitem[]{} Allen, L. E., Myers, P. C., Di Francesco, J., Mathieu, R., Chen, H., \& Young, E. 2002, \apj, 566 993
\bibitem[]{} Andr\'{e}, P., Mart\'{i}n-Pintado, J., Despois, D., \& Montmerle, T. 1990, A\&A, 236, 180
\bibitem[]{} Andr\'{e}, P., Ward-Thompson, D., \& Barsony, M. 1993, \apj, 406, 122
\bibitem[]{} Andr\'{e}, P., \& Montmerle, T. 1994, \apj, 420, 837
\bibitem[]{} Andr\'{e}, P., Ward-Thompson, D., \& Barsony, M. 2000, in Protostars and Planets IV, ed. V. Mannings, A. Boss, S. Russell. (Tucson: Univ. Arizona Press), 61
\bibitem[]{} Andr\'{e}, P., Basu, S., \& Inutsuka, S. 2008, preprint (astro-ph/0801.4210)
\bibitem[]{} Arce, H. G., Shepherd, D., Gueth, F., Lee, C., Bachiller, R., Rosen, A., \& Beuther, H. 2007, in Protostars and Planets V, ed. B. Reipurth, D. Jewitt, K. Keil. (Tucson: Univ. Arizona Press)
\bibitem[]{} Arce, H. G., \& Sargent, A. I. 2006, \apj, 646, 1070
\bibitem[]{} Avery, L. W., Hayashi, S. S., \& White, G. J. 1990, \apj, 357, 524
\bibitem[]{} Bachiller, R. 1996, ARA\&A, 34, 111
\bibitem[]{} Bacmann, A., Lefloch, B., Ceccarelli, C., Castets, A., Steinacker, J., \& Loinard, L. 2002, A\&A, 389, L6
\bibitem[]{} Barsony, M., Kenyon, S. J., Lada, E. A., \& Teuben, P. J. 1997, ApJS, 112 109
\bibitem[]{} Bontemps, S., Andr\'{e}, P., Terebey, S., \& Cabrit, S. 1996, A\&A, 311, 858
\bibitem[]{} Bontemps, S., et al. 2001, A\&A, 372, 173 
\bibitem[]{} Buckle, J.V., Curtis, E., Dent, W., Hills, R., \& Richer, J. 2006, in IAU Symposium No. 237, Triggered Star Formation in a Turbulent ISM, ed. B. G. Elmegreen, J. Palous (Prague: IAUS), 87
\bibitem[]{} Carrasco-Gonz\'{a}lez, C., Anglada, G., Rodr\'{i}guez, L. F., Torrelles, J. M., Osorio, M., \& Girart, J. M. 2008, \apj, 676, 1073 
\bibitem[]{} Caselli, P., Walmsley, C. M., Tafalla, M., Dore, L., \& Myers, P.C. 1999, \apj, 523, L165
\bibitem[]{} Clube, S. V. M. 1967, MNRAS, 137, 189
\bibitem[]{} Comeron, F., Torra, J., \& Gomez, A. E. 1992, Ap\&SS, 187, 187
\bibitem[]{} Crapsi, A., Caselli, P., Walmsley, C. M., Tafalla, M., Lee, C. W., Bourke, T. L., \& Myers, P. C. 2004, A\&A, 420, 957
\bibitem[]{} Elias, J. H. 1978, \apj, 224, 453
\bibitem[]{} Evans, N. J., II, et al. 2003, PASP, 115, 965
\bibitem[]{} Grasdalen, G. L., Strom, K. M., \& Strom, S. E. 1973, \apj, 184, L53
\bibitem[]{} Gregersen, E. M., Evans, N. J., II, Zhou, S., \& Choi, M. 1997, \apj, 484, 256 
\bibitem[]{} Gregersen, E.M., \& Evans, N.J., II, 2000, \apj, 538, 260
\bibitem[]{} Gregersen, E. M., Evans, N. J., II, Mardones, D., \& Myers, P. C., 2000, \apj, 533, 440
\bibitem[]{} Johnstone, D., Wilson, C. D., Moriarty-Schieven, G., Joncas, G., Smith, G., Gregersen, E., \& Fich, M. 2000, \apj, 545, 327
\bibitem[]{} Johnstone, D., Fich, M., Mitchell, G. F., \& Moriarty-Schieven, G. 2001, \apj, 559, 307
\bibitem[]{} Johnstone, D., Di Francesco, J., Kirk, H. 2004, \apj, 611, L45
\bibitem[]{} J\o rgensen, J. K., Sch\"{o}ier, F. L., \& van Dishoeck, E. F. 2005, A\&A, 435, 177
\bibitem[]{} J\o rgensen, J. K., Johnstone, D., Kirk, H., \& Myers, P. C. 2007, \apj, 656, 293
\bibitem[]{} J\o rgensen, J. K., Johnstone, D., Kirk, H., Myers, P. C., Allen, L. E., \& Shirley, Y. L. 2008, ApJ, in press (astro-ph/0805.0599)
\bibitem[]{} Klapper, G., Surin, L., Lewen, F., \&  M\"{u}ller, H.S.P., Pak, I., \& Winnewisser, G. 2003, \apj, 582, 262
\bibitem[]{} Klaassen, P. D., Plume, R., Ouyed, R., von benda-Beckmann, \& Di Francesco, J. 2006, ApJ, 648, 1079
\bibitem[]{} K\"{o}nigl, A., \& Pudritz, R. E. 2000, in Protostars and Planets IV, ed. V. Mannings, A. Boss, S. Russell. (Tucson: Univ. Arizona Press),  760
\bibitem[]{} Kroupa, P. 2001, in ASP Conf. Ser. 228, Dynamics of Star Clusters and the Milky Way, ed. S. Deiters, B. Fuchs, A. Just, R. Spurzem, \& R. Wielen (San Francisco: ASP), 187
\bibitem[]{} Kutner, M. L., \& Ulich, B. L. 1981, \apj , 250, 341
\bibitem[]{} Lada, C. J. 1985, ARA\&A, 23, 267
\bibitem[]{} Lada, C. J., Muench, A. A., Rathborne, J., Alves, J. F., \& Lombardi, M. 2008, \apj, 672, 410
\bibitem[]{} Lee, C. W., Myers, P. C., \& Tafalla, M. 1999, \apj, 526, 788
\bibitem[]{} Leung, C. M., \& Brown, R. L. 1977, \apj, 214, L73
\bibitem[]{} Mardones, D., Myers, P. C., Tafalla, M., Wilner, D. J., Bachiller, R., \& Garay, G. 1997, \apj, 489, 719
\bibitem[]{} Mizuno, A., Fukui, Y., Iwata, T., Nozawa, S., \& Takano, T. 1990, \apj, 356, 184
\bibitem[]{} Motte, F., Andr\'{e}, P., \& Neri, R. 1998, A\&A, 336, 150
\bibitem[]{} Myers, P. C., Mardones, D., Tafalla, M., Williams, J. P., \& Wilner, D. J., 1996, \apj, 465, L133
\bibitem[]{} Richer, J. S., Shepherd, D. S., Cabrit, S., Bachiller, R., \& Churchwell, E. 2000, in Protostars and Planets IV, ed. V. Mannings, A. Boss, S. Russell. (Tucson: Univ. Arizona Press), 869
\bibitem[]{} Salpeter, E. E. 1955, \apj, 121, 161 
\bibitem[]{} Smith, H., et al. 2003, in Proceedings of the SPIE 4855, Millimeter and Submillimeter Detectors for Astronomy, ed. T. G. Phillips, J. Zmuidzinas (Waikoloa, HI: SPIE), 338
\bibitem[]{} Sohn, J., Lee, C.W., Park, Y., Lee, H. M., Myers, P. C., \& Lee, Y. 2007, \apj, 664, 928
\bibitem[]{} Stothers, R., \& Frogel, J. A. 1974, AJ, 79, 456
\bibitem[]{} Testi, L., \& Sargent, A. I. 1998, \apj, 508, L91
\bibitem[]{} Tobin, J. J., Looney, L. W., Mundy, L. G., Kwon, W., \& Hamidouche, M. 2007, ApJ, 659, 1404
\bibitem[]{} Vrba, F. J., Strom, K. M., Strom, S. E., \& Grasdalen, G. L. 1975, \apj, 197, 77
\bibitem[]{} Wannier, P. G. 1980, ARA\&A, 18, 399
\bibitem[]{} Ward-Thompson, D. et al. 2007, PASP, 119, 855
\bibitem[]{} Werner, M. W., et al. 2004, ApJS, 154, 1
\bibitem[]{} Willacy, K., Langer, W. D., \& Velusamy, T. 1998, \apj, 507, L171
\bibitem[]{} Young, E. T., Lada, C. J., \& Wilking, B. A. 1986, \apj, 304, L45
\bibitem[]{} Young, K. E., et al. 2006, \apj, 644, 326


\end{thebibliography}
 \end{document}